\documentclass[a4paper,12pt]{article}
\usepackage{amssymb}
\usepackage{amsfonts, amsmath}
\usepackage{natbib}
\usepackage{comment}
\usepackage{hyperref}

\newtheorem{theorem}{Theorem}

\newtheorem{example}{Example}

\newtheorem{lemma}{Lemma}

\newtheorem{proposition}{Proposition}

\newenvironment{proof}[1][Proof]{\noindent\textbf{#1.} }{\ \rule{0.5em}{0.5em}}

\hypersetup{
    colorlinks=true,
    citecolor=blue,
    linkcolor=blue,
    filecolor=blue,
    urlcolor=blue,
    pdftitle={Moral Consumers and the Environment},
    pdfpagemode=FullScreen,
    }
\begin{document}

\title{{Moral Consumers and the Environment}\thanks{%
We thank Ingela Alger, Geir Asheim, Roland Bénabou, Kjell Arne Brekke, Armin Falk, Bård Harstad, Frikk Nesje, Klaus Schmidt, and seminar participants in
Copenhagen, Paris, London, and Lille, as well as participants in the Cornell-Bath Conference on Technology and Labour held at the University of Bath, for many generous comments. We have worked on the paper for several years. Recently, we have begun using Claude for assistance, which has led us to catch a few mathematical bugs.} }
\author{Mogens Fosgerau \\
{\small University of Copenhagen}
\and J\"{o}rgen Weibull \\
{\small Stockholm School of Economics} }
\date{ \today }
\maketitle
\begin{abstract}
We study the taxation of environmentally harmful goods when consumers have a Kantian desire to ``do the right thing''. A consumer evaluates her buying strategy, a rule that specifies whether to buy, by the material payoff she would obtain if others followed the same rule with a probability given by her degree of morality. Equilibrium exists and is unique in finite populations with heterogeneous morality. The optimal tax depends crucially on whether consumers are aware that tax revenue is returned to them. If they are, the Pigovian tax is optimal under homogeneous morality, whatever consumers know about environmental costs, and we give conditions under which it remains the unique optimum under heterogeneous morality. If they are not, the optimal tax is lower, and more so the more moral consumers are. Under a quota system such as the EU Emissions Trading System, any cap at least as stringent as the Pigovian quota admits an equilibrium in which morality has no effect on consumers who are aware of the system.
\end{abstract}

\noindent \textbf{Keywords}: climate, morality, environment, optimal taxation.

\noindent \textbf{JEL codes}: H23, H21, Q58, D11, D62 

\section{Introduction}

Standard economic analysis of environmental policy is premised on the assumption that consumers are selfish. As a result, it is optimal to tax each unit of consumption at its environmental cost \citep{pigou_economics_1920}. Yet many people feel morally obliged to reduce their consumption of environmentally harmful goods \citep{european_investment_bank_eib_2022}. A growing literature documents and models deontological motives (a desire to ``do the right thing'') alongside self-interest \citep{van_leeuwen_estimating_2024,ellingsen_model_2026,benabou_ends_2024}.

This tension motivates the question we address: how should one tax a consumer population that differs not only in its valuations of goods but also in its degrees of morality? Fundamentally, our aim is to help improve the economic analysis of environmental policy by engaging with the moral motivations that people arguably have. More broadly, we hope that taking such moral concerns seriously will help economists to have more meaningful and less emotionally charged conversations with non-economists.

We model moral motivations in the Kantian spirit of \citet{laffont_macroeconomic_1975} by using the \emph{homo moralis} preferences of \citet{alger_homo_2013,alger_evolution_2016,alger_evolutionary_2019}. These preferences capture a deontological motive, distinct from altruism and social norms \citep{nyborg_karine_social_2018}. A \emph{homo moralis} consumer has preferences over strategies: rules that specify whether to buy a good given its price, its hedonic value to the consumer, and information about its environmental cost. She evaluates such strategies by the material payoff she would obtain if a share $\kappa$ of all other consumers were to use the same strategy. With $\kappa =0$, she is a \emph{homo economicus} who is concerned only with the consequences for herself of her own actions; with $\kappa =1$, she is a \emph{homo kantiensis}, who follows Kant's categorical imperative \citep{kant_groundwork_1785}: ``Act only according to that maxim whereby you can at the same time will that it should become a universal law.'' In general, $0 \leq \kappa\leq 1$ expresses her degree of Kantian morality. It is a preference parameter, not a belief: she does not expect others to imitate her. Instead, she evaluates strategies by a thought experiment: what would happen if others were to act like her? \citet{alger_homo_2013,alger_evolution_2016} derive these preferences from evolutionary stability when preferences are private information, with a positive degree of morality arising under assortative matching.\footnote{We do not study the evolution of preferences and moral values in the product market considered here, but take consumers' degrees of morality as given.}

We embed \emph{homo moralis} consumers in a simple market where each consumer decides whether to buy a unit of a good belonging to a given category of goods. The environmental cost of a unit, that is, the social cost of the environmental damage caused by its production and/or consumption, is shared equally by everyone in the population. Consumers differ in their valuation of the unit, in what they know about its environmental cost, and in their degree of morality. A planner who knows the environmental cost of each unit and the distribution of valuations, information and morality, but not individual valuations or degrees of morality, taxes each unit and returns the revenue to all consumers as a lump-sum transfer. The planner maximizes consumers' utility, including its moral component, evaluated at realized environmental costs, taxes and transfers.

We use this framework to ask whether environmental taxes should be lower (or higher) than the Pigovian tax when consumers are moral. 
It turns out that the answer depends crucially on whether consumers are aware of the tax-transfer system.\footnote{The role of awareness relates our analysis to behavioral public finance, where consumers may misperceive taxes \citep{chetty_salience_2009,farhi_optimal_2020}. Our consumers perceive prices correctly; what matters is whether they know what happens to tax revenue.}
In a large population, we find that a consumer with degree of morality $\kappa$ buys the unit if and only if her consumer surplus, that is, her hedonic valuation of the unit minus its price, exceeds the share $\kappa$ of the environmental cost that she expects the unit to cause or, if she is aware that the tax revenue is returned to consumers, the share $\kappa$ of that expected cost net of the tax revenue that she expects the purchase of the unit to generate.

Suppose first that all consumers are equally moral, but not pure Kantians. If they are aware of the tax-transfer system, the Pigovian tax is the unique optimal tax, whatever their degree of morality and whatever their information about the environmental cost. The reason is that under the Pigovian tax, the tax revenue that accrues to the consumer population is exactly equal to the environmental cost that it bears. Consumers, irrespective of their degree of morality, then behave as if they were \emph{homo economicus}. This is not a case of moral behavior being crowded out \citep{ostrom_crowding_2000}; consumers remain moral, but Pigovian taxation makes the net external cost of a purchase equal to zero. In this case, information about environmental costs, for example via eco-labels, has no effect on behavior, since the price already incorporates the planner's information about the environmental cost.

Suppose now that consumers are unaware of the return of revenue. Then only the expected environmental cost enters their moral calculus, and the optimal tax is lower than the Pigovian tax. If consumers have perfect information about the environmental cost, the optimal tax is $\left(1-\kappa\right)z$, where $z$ is the environmental cost of the unit.\footnote{\citet{eichner_kantians_2022} obtain the same result for identical moral consumers who do not see through the government budget constraint.} With no information about environmental costs, the optimal tax is $z-\kappa\bar{z}$, where $\bar{z}$ is the average environmental cost. This makes the tax in fact a subsidy for goods whose environmental cost is below $\kappa\bar{z}$. Uninformed moral consumers, who attribute the average environmental cost to every good, would otherwise buy too little of such goods.

Under homogeneous morality, the optimal tax thus implements the first best when consumers are aware, whatever their information, and when they are unaware and either perfectly informed or uninformed. By contrast, in the more realistic case of heterogeneous morality and unaware consumers, a single tax cannot align everyone's behavior with the first best, and any optimal tax is below the Pigovian tax on goods that consumers expect to be harmful.
For aware consumers, the Pigovian tax satisfies the first-order conditions for optimality for any distribution of morality, and we give conditions under which it is the unique optimum.

In reality, some consumers are likely to be aware and others not. Survey evidence from Canada and Switzerland shows that many residents were unaware of the carbon-tax rebates they received \citep{mildenberger_limited_2022}. With mixed awareness, no single tax can align the behavior of the two groups. For perfectly informed consumers with a common degree of morality, we show that welfare then falls short of the first best. Making the tax-transfer system transparent is therefore valuable.

Instead of a tax, consider a cap on emissions implemented through tradable allowances, as in the European Union Emissions Trading System \citep{bayer_european_2020}, and suppose that the allowance revenue is returned to consumers as a uniform lump-sum transfer. Consumers who are unaware of the quota system respond to the allowance price as they would respond to a tax.
Consumers who are aware of it reason differently. 
Under a binding cap, total emissions are fixed, so a moral consumer who asks what would happen if others adopted her strategy finds that additional purchases would only raise the allowance price, not emissions.
We show that if the cap is sufficiently tight, there is an equilibrium in which aware consumers behave as \emph{homo economicus}, whatever their information and whatever their degree of morality (below one), and the optimal such cap reproduces the allocation under Pigovian taxation. Under a tax, by contrast, aware moral consumers respond to the difference between the environmental cost and the tax. Morality thus introduces a distinction between price and quantity regulation that does not depend on the regulatory uncertainty emphasized by \citet{weitzman_prices_1974}. \citet{herweg_how_2022} find a related offsetting mechanism in a model where consumers suffer a moral cost if their consumption of a dirty good exceeds a social norm.

In contrast to models of warm glow \citep{andreoni_giving_1989,andreoni_impure_1990}, where the moral payoff is attached to the act itself, \emph{homo moralis} cares about the material consequences that would follow if others used her strategy. Her moral concern therefore depends on environmental policy, which is why awareness and the choice of policy instrument matter in our model. See Section \ref{sec:related literature} for a more detailed discussion.

The paper proceeds as follows. Section~\ref{sec:model} presents the model, Section~\ref{sec:optimal taxation} analyzes optimal taxation, Section~\ref{sec:compstat} presents comparative statics, and Section~\ref{sec: quota system} analyzes quotas. Section~\ref{sec:extensions} contains extensions, including mixed awareness, Section~\ref{sec:related literature} discusses related literature, and Section~\ref{sec:discussion} concludes. Proofs are in the Appendix.

\section{Model\label{sec:model}}

\subsection{Consumers and goods}

To enable a precise game-theoretic analysis of morality, we set up a model for a finite population of consumers $i\in I=\left\{ 1,...,n\right\} $. Each
consumer considers whether to buy a unit of a good in some category of goods that entail environmental harm. The decision of each consumer $
i$ is a binary variable $b_{i}=1$ (buy) or $b_{i}=0$ (not buy). Aggregate consumption of goods in the
category is thus $\sum_{i\in I}b_{i}$.

At the moment of purchase, the individual consumer has a subjective hedonic
valuation $v_{i}$ of the good at hand, where $v_{i}$ is a realization of a
random variable $V_{i}$. Valuations are statistically independent and
identically distributed according to a CDF $F$ that has a continuous, differentiable and everywhere
positive PDF $f$ and finite mean, $\mathbb{E}[V]<\infty$. Let $Z_{i}$ be the random variable that represents the
environmental cost to society associated with the production and/or
consumption of the unit that consumer $i$ considers for purchase. Thus every
individual $j\in I$ (including $i$ herself) will experience the
environmental cost or disutility $Z_{i}/n$ if $i\in I$ buys the unit in
question. Aggregate consumption therefore entails an environmental cost or
disutility $\sum_{i}b_{i}Z_{i}/n$ for each individual $j$ in the population.
Valuations $V_{i}$ and environmental costs $Z_{i}$ are statistically
independent, also across consumers.

The price of the unit of the good considered by consumer $i$ is $%
P_{i}=\omega \left( Z_{i}\right) +\tau \left( Z_{i}\right) $, where $\omega
\left( Z_{i}\right) >0$ is its production cost and $\tau \left(
Z_{i}\right) \ $the environmental tax, chosen by a social planner who knows
the environmental cost associated with any unit of the good and designs the tax schedule $\tau :\mathbb{R}_{+}\rightarrow \mathbb{R}$
accordingly. There is perfect competition among sellers, so these earn zero
profit. A consumer $i$ who considers purchase of a unit observes its
realized price $p=\omega \left( z\right) +\tau \left( z\right) $, where $z$
is the realization of $Z_{i}$ for the unit in question. Consumers do not
observe $\omega \left( z\right) $ or $\tau \left( z\right) $, only their
sum, and they do not know the function $\omega \left( \cdot \right) $. They
may have some information about the environmental cost of the good at hand,
and they may or may not know the tax schedule $\tau \left( \cdot \right) $.
Each of these cases will be analyzed separately. At one extreme, consumers
observe the exact environmental cost of the unit at hand, the case of \emph{%
perfect information}, and they know the tax schedule (they are\ \emph{tax
aware}). At the opposite extreme, consumers have \emph{no information} about
the environmental cost of the unit at hand, they only know the average cost,
and they do not know the tax schedule (they are \emph{tax unaware}).\footnote{Awareness is here binary and universal. In practice it is partial and
heterogeneous across consumers. Section \ref{sec:ext:awareness} considers a
population in which only some consumers are aware, and shows that this is
not innocuous: no single tax then aligns the behavior of both groups.
Intermediate degrees of awareness are formally possible but lack a
clear interpretation, and we do not pursue them.}

Tax revenues are paid back to the consumer population. Every individual thus
receives the lump-sum transfer
\begin{equation}
T=\frac{1}{n}\sum_{j\in I}b_{j}\tau \left( Z_{j}\right) ,  \label{eq:T}
\end{equation}%
where $Z_{j}$ is the environmental cost of $j$'s purchase (if made).
Consumers who are aware of the tax schedule $\tau (\cdot )$ are also aware
of how the per capita transfer $T$ depends on aggregate consumption through %
\eqref{eq:T}. Unaware consumers, those who are unaware of the tax schedule,
are also unaware of the transfer. They only see the consumer price and
expect no transfer. We model this difference in awareness by including the
term $\alpha T$ in the consumer utility, where $\alpha $ is deterministic
and common to all consumers, with $\alpha =1$ meaning that they all are
aware of the tax-transfer scheme and $\alpha =0$ that they are unaware of it.

In sum, the model allows for a continuum of information scenarios but only two awareness scenarios. Several of our results hold for arbitrary signals. Others are established only for the two boundary cases of perfect and no information, which, combined with the two awareness scenarios, give four boundary cases.

\subsection{Information about environmental costs}\label{sec:information}

As mentioned above, we will study the role of information about
environmental costs in consumer decisions. We do this by assuming that each
buyer observes a private signal about the environmental cost of the unit she
considers. For each consumer $i\in I$, the signal $S_{i}$ may be informative
about the environmental cost $Z_{i}$ associated with the unit considered by $%
i$, but $S_{i}$ is statistically independent of the consumer's valuation $%
V_{i}$. The joint distribution of $(Z_{i},S_{i})$ is i.i.d. across
consumers. For analytical convenience, we assume that this joint
distribution has finite support. Having observed her signal $S_{i}$,
consumer $i$ forms the conditional expectation $\hat{Z}_{i}=\mathbb{E}\left[
Z_{i}\mid S_{i}\right] $, a random variable with expectation $\mathbb{E}%
\left[ \hat{Z}_{i}\right] =\bar{z}$ for all $i\in I$, where $\bar z = \mathbb E[Z]$. We pay special attention to the boundary cases of \emph{perfect information}, $S_{i}=Z_{i}$, so that $\hat{Z}_{i}=Z_{i}$, and \emph{no information}, $S_{i}\equiv \bar{z}$, so that $\hat{Z}_{i}\equiv \bar{z}$.

An aware consumer $i$ who knows the tax-transfer system and receives a signal $S_{i}$ about its environmental cost, forms a conditional expectation of her own contribution $R_{i}=\tau (Z_{i})$ to the social planner's tax revenues in case the consumer buys a unit of the good. We write this conditional expectation as $\hat{R}_{i}=\mathbb{E}[{\tau (Z_{i})}\mid S_{i}]$.

Consumers observe neither $\omega(Z_i)$ nor $\tau(Z_i)$ separately and do not
know the function $\omega(\cdot)$. We therefore assume that they regard the
price as uninformative about the environmental cost, given their signal.
Formally, the expectations in \eqref{pii}--\eqref{U cond} are taken with
respect to beliefs under which $P_i$ is independent of $Z_i$ conditionally on
$S_i$.

\subsection{Hedonic utility}\label{sec:hedonic}

The perceived \textit{hedonic} \textit{utility} to consumer $i\in I$ is the additively
separable random variable
\begin{equation}\label{ui}
U_{i}=\left( V_{i}-P_{i}\right) b_{i}-\frac{1}{n}\sum_{j\in I}\left(
Z_{j}-\alpha R_{j}\right) b_{j}
\end{equation}%
The first term is the usual consumer surplus --- valuation minus price ---
that the individual obtains from consuming a unit of the good. The second
term is the difference between the disutility that the consumer experiences from the environmental
harm caused by aggregate consumption of the goods in the category, and the
transfer that will arrive---and which in part depends on
her tax payment if she buys the good. If consumers are unaware of the
tax-transfer system ($\alpha =0$) then the term $R_{i}$ vanishes from their
perceived hedonic utility as they make their buying decisions.

We note that a \emph{homo economicus} consumer, having  observed the (realized) price of the unit at hand, learned her valuation of it and received her signal about its environmental cost, buys the unit if and
only if
\begin{equation*}
V_{i}-P_{i}>\frac{1}{n}\left( \hat{Z}_{i}-\alpha \hat{R}_{i}\right) ,
\end{equation*}%
that is, if her consumer surplus exceeds the environmental harm she expects to cause herself, net of the expected increase in her transfer due to her purchase in case she is aware ($\alpha=1$), and conditional
upon the signal she has received about its environmental cost.

A morally concerned consumer's buying decision is more complex; she also considers the harm and/or benefit her purchase may cause others, and contemplates what is ``the morally right thing" to do. In order to deal with this, we use the \emph{homo moralis} paradigm of \citet{alger_homo_2013,alger_evolution_2016}. In order to apply it, one needs to first cast the decisions made by all consumers as strategy choices in an \textit{ex ante} symmetric game in ``material", here hedonic, payoffs.

Let the production cost function $\omega $ and the tax function $\tau $ be given, along with the CDF $F$ of valuations and the joint distribution of environmental costs and signals $\left( Z_{i},S_{i}\right) $. In the associated game, a (pure) \textit{strategy} for player $i$ is a (Borel measurable) function $x_{i}:\mathbb{R}_+ \times \mathbb{R} \times \operatorname{supp}\left( S\right)
\rightarrow \{0,1\}$ that specifies a buying decision $%
b_{i}=x_{i}\left( v_{i},p_{i},s_{i}\right) $ for any given realizations of
own valuation $V_{i}$, price $P_{i}$, and signal $S_{i}$. Denote the common
strategy space $X$. The payoff to a consumer/player in this game is the
consumer's \textit{ex ante} (that is, before the price, valuation and signal
are known) \textit{expected hedonic utility} (given his or her awareness).
This is a real-valued function on $X^{n}$, whose values depend on her own
strategy, $x_{i}$, and on the strategy vector $x_{-i}=\left( x_{j}\right)
_{j\neq i}$, of all other consumers:
\begin{eqnarray}
\pi \left( x_{i},x_{-i}\right) &=&\mathbb{E}\left[ \left( V_{i}-P_{i}-\frac{%
Z_{i}-\alpha R_{i}}{n}\right) x_{i}\left( V_{i},P_{i},S_{i}\right) \right]
\label{pii} \\
&&-\mathbb{E}\left[ \sum_{j\neq i}\frac{Z_{j}-\alpha R_{j}}{n}x_{j}\left(
V_{j},P_{j},S_{j}\right) \right] ,  \notag
\end{eqnarray}%
where the first expectation concerns everything that depends on her own
``type" $\left( V_{i},P_{i},S_{i}\right) $ and strategy $x_{i}$, and the
second expectation concerns everything that depends on others' types $\left(
V_{j},P_{j},S_{j}\right) $ and strategies $x_{j}$ (for $j\neq i$).

Equation \eqref{pii} shows that the game is aggregative; the payoff to a
consumer/player is invariant under permutation of other consumers'/players'
strategies. All that matters is the total environmental cost and transfer that they cause.

\subsection{Moral utility}

The game-theoretic formulation of consumers' decision problems just given
allows us to apply the \emph{homo moralis} utility function. The \textit{%
ex ante} utility of a \emph{homo moralis} consumer is the
(material/hedonic) payoff that she would obtain in a hypothetical scenario
where some of the other consumers were also to use her strategy. This
embodies a generalization of the deontological Kantian notion to do ``what
would be right to do if \textit{all} others were to act like you" to do
``what would be the right thing to do if \textit{some} others were to act
like you". The degree of morality is a number between zero and one and
represents the probability assigned to others acting like you.

It is not the case that such a moral decision-maker (falsely) believes that
others will act like her. It is merely a way to operationally express her
preferences over strategy profiles. She ranks these based on what would
happen if, hypothetically, others were, with some probability, to also use
her strategy.

More precisely, the utility to a \emph{homo moralis} consumer $i\in I$
with degree of morality $\kappa _{i}\in \left[ 0,1\right] $ is the expected
(material/hedonic) payoff associated with the scenario in which every other
consumer $j\neq i$, with probability $\kappa _{i}$ independently of each
other switches to consumer $i$'s strategy. In particular, \emph{homo
moralis} with morality $\kappa _{i}=0$ has the preferences of a \emph{homo
economicus}. In the opposite extreme case, when $\kappa _{i}=1$, \emph{homo moralis} has purely Kantian preferences, that is, she chooses a
strategy that would be optimal in terms of her material/hedonic payoff if
\textit{all} others were to use that same strategy.

The ex ante utility function $\tilde{U}_{i}:{X}^{n}\rightarrow \mathbb{R}$
of a \emph{homo moralis} consumer $i$ with degree of morality $\kappa _{i}\in %
\left[ 0,1\right] $ interacting in a symmetric $n$-player game with strategy
space ${X}$ and material/hedonic payoff function $\pi :X^{n}\rightarrow
\mathbb{R}$, is defined for all strategy profiles $\left(
x_{i},x_{-i}\right) \in {X}^{n}$ as follows. Let $K_{ij},i,j\in I,i\neq j$
be i.i.d. Bernoulli random variables that take the value $1$ with
probability $\kappa _{i}$ and the value $0$ with probability $1-\kappa _{i}$%
. These random variables control who will be assumed, in $i$'s consideration
of her consumption decision, to use the same strategy as herself:%
\begin{eqnarray}
\tilde{U}_{i}\left( x_{i},x_{-i}\right) &=&\mathbb{E}\left[ \left(
V_{i}-P_{i}-\frac{Z_{i}-\alpha R_{i}}{n}\right) x_{i}\left(
V_{i},P_{i},S_{i}\right) \right]  \label{ES U} \\
&&-\sum\limits_{j\neq i}\mathbb{E}\left[ \frac{Z_{j}-\alpha R_{j}}{n}%
K_{ij}x_{i}\left( V_{j},P_{j},S_{j}\right) \right]  \notag \\
&&-\sum\limits_{j\neq i}\mathbb{E}\left[ \frac{Z_{j}-\alpha R_{j}}{n}\left(
1-K_{ij}\right) x_{j}\left( V_{j},P_{j},S_{j}\right) \right] .  \notag
\end{eqnarray}

In other words, consumer $i$, before learning her valuation of the unit at
hand, its price and her signal, evaluates her strategy $x_{i}\in X$ in the
light of what would happen if others would switch from their current
strategy $x_{j}\in X$ ($j\neq i$) to her strategy with probability $\kappa
_{i}$. If the consumer has degree of morality $\kappa _{i}=0$ she is a \emph{homo
economicus}, and does not at all consider the consequences if others were to
act like her. At the other extreme, when $\kappa _{i}=1$, the consumer is a pure Kantian moralist, and acts according to the maxim (here strategy) that she would like all consumers to use.

\subsection{Optimal consumer strategies}

Maximizing the partly morally motivated utility function in (\ref{ES U}),
for given strategies by other consumers, leads to the consumer's optimal
strategy (or strategies), a decision rule that for every realization of $%
\left( V_{i},P_{i},S_{i}\right) $ maximizes the associated conditionally
expected utility, conditioned upon the consumer's signal. Hence, the
expected utility that the strategy $x_{i}$ has to maximize is {\normalsize
\begin{eqnarray}
\tilde{U}_{i}\left( x_{i},x_{-i}\right) &=&\mathbb{E}\left[ \left(
V_{i}-P_{i}-\left( \frac{1}{n}+\frac{n-1}{n}\kappa _{i}\right) \left( \hat{Z}%
_{i}-\alpha \hat{R}_{i}\right) \right) x_{i}\left( V_{i},P_{i},S_{i}\right) %
\right]  \notag \\
&&-\frac{1-\kappa _{i}}{n}\sum\limits_{j\neq i}\mathbb{E}\left[ \left( \hat{Z%
}_{j}-\alpha \hat{R}_{j}\right) x_{j}\left( V_{j},P_{j},S_{j}\right) \right]
,  \label{U cond}
\end{eqnarray}%
}
where $\hat{Z}_i=\mathbb{E}[Z_i\mid S_i]$ and $\hat{R}_i=\mathbb{E}[\tau(Z_i)\mid S_i]$ are the conditional expectations introduced in Section~\ref{sec:information}.

Because of the additive separability of the consumers utility and the
linearity of the environmental cost, optimality of a consumer's strategy is
independent of what strategies other consumers use. For optimality it is
necessary and sufficient that the consumer maximizes what he or she can
influence, that is, the first expectation. Let $\tilde{\kappa}_i:=\frac{1}{n}+\frac{n-1}{n}\kappa_i$ denote consumer $i$'s effective degree of morality, the weight she attaches to the environmental term in \eqref{U cond}. Then:

\begin{lemma}
\label{opt x}An optimal strategy for consumer $i\in I$, for any given
strategy profile $x_{-i}$, is
\begin{equation}
x_{i}^{\ast }\left( V_{i},P_{i},S_{i}\right) =\left\{
\begin{array}{ll}
1 & \text{if }V_{i}-P_{i}>\tilde \kappa_i
\left( \hat{Z}_{i}-\alpha \hat{R}_{i}\right) \\
0 & \text{otherwise}%
\end{array}%
\right.  \label{x*}
\end{equation}
\end{lemma}

We first note that the optimal strategy is independent of what strategies other consumers use. Hence, it is a dominant strategy. Second, we note that the term on the right-hand side of the inequality is a threshold that the
hedonic consumer surplus must exceed for the partly morally motivated
consumer to buy the good.\footnote{%
The consumer is indifferent in case of equality in (\ref{x*}). However,
these situations will not matter in the subsequent analysis since they occur
with probability zero.} The threshold is always non-negative for unaware
consumers ($\alpha =0$). The threshold is also positive for an aware
consumer who expects the environmental cost to exceed the transfer.
Unsurprisingly, in these two cases, the threshold increases with the
consumer's morality; more moral consumers are less willing to buy environmentally harmful goods.

We note that a maximally moral consumer, a \emph{homo kantiensis} with
degree of morality $\kappa _{i}=1$, buys the good if and only if
\begin{equation*}
V_{i}-P_{i}>\hat{Z}_{i}-\alpha \hat{R}_{i},
\end{equation*}%
(equality occurs with probability zero and is disregarded). If, furthermore,
\emph{homo kantiensis} is aware of the transfer, $\alpha =1$, and the tax is
Pigovian, i.e., $\tau (Z_{i})=Z_{i}$, then $\hat{R}_{i}=\hat{Z}_{i}$ and
thus this inequality reduces to $V_{i}>\omega \left( Z_{i}\right) +{Z}_{i}$.
This means that such a \emph{homo kantiensis} buys a unit if and only if
her hedonic valuation exceeds its production cost plus its environmental
cost. At the opposite end of the morality spectrum, $\kappa _{i}=0$, we have
\emph{homo economicus}, who only considers the material/hedonic
consequences of her own action for herself. Such a consumer buys the good if
and only if
\begin{equation*}
V_{i}-P_{i}>\frac{1}{n}\left( \hat{Z}_{i}-\alpha \hat{R}_{i}\right) .
\end{equation*}%
That is, her threshold is $1/n$ times the size of \emph{homo kantiensis}'
threshold. The economicus threshold is close to zero when the consumer
population is large and tends to zero as $n\rightarrow \infty $. In an
infinite population, \emph{homo economicus} buys a unit if and only if the hedonic
surplus is positive for her.

Moral consumers thus forgo some purchases they would hedonistically value,
and thereby obtain lower expected hedonistic utility/payoff than purely
selfish consumers. This utility loss, or \textquotedblleft cost of being
moral,\textquotedblright\ occurs precisely when a consumer's valuation
exceeds the price but falls short of her threshold for buying, given in
equation (\ref{x*}).

\subsection{Consumer equilibrium}

We view the interaction between consumers as a simultaneous-move game under
incomplete information, where each consumer's valuation and degree of
morality are private information. Existence of Nash equilibrium is
guaranteed, since by Lemma \ref{opt x} every consumer has a dominant strategy, so these together constitute a Nash equilibrium. Considering uniqueness, eq. \eqref{ES U} shows that a consumer's strategy may be altered on a (Borel measurable) null set of valuations for each signal value without affecting
the consumer's expected utility. Hence, each consumer has infinitely many optimal strategies, but they all agree almost surely with (\ref{x*}). In sum:

\begin{proposition}\label{prop:equilibrium}
There exists a Nash equilibrium in the game between $n$ \emph{homo moralis} consumers of arbitrary degrees of morality. The equilibrium is uniquely determined by \eqref{x*} for every consumer $i$, every signal value, and almost all valuations. Expected equilibrium utilities are uniquely determined by %
\eqref{ES U}.
\end{proposition}

With additive utility and linear environmental costs, the equilibrium involves no real interaction between consumers. In particular, they need not form expectations about others' strategies to find their own optimal strategy. This is no longer so when environmental costs are non-linear, see Section \ref{sec:ext:convex}.

\section{Optimal taxation\label{sec:optimal taxation}}


Consider a benevolent social planner who taxes consumption of environmentally harmful goods in order to maximize consumers' utility, evaluated at realized outcomes. This differs from the expected utility that consumers maximize, given their information at the time of purchase, in that the difference $\hat{Z}_{i}-\alpha \hat{R}_{i}$ between the consumer's conditionally expected environmental cost and transfer of tax revenues in the integrand in \eqref{U cond} is replaced by the actual difference $Z_{i}-\tau (Z_{i})$. Consumer $i$'s ex ante expected utility at
realized outcomes, of relevance for the social planner, is thus

\begin{eqnarray}
U_{i}^{\ast }\left( x_{i},x_{-i}\right) &=&\mathbb{E}\left[ \left(
V_{i}-P_{i}-\tilde \kappa_i \left(
Z_{i}-\tau \left( Z_{i}\right) \right) \right) x_{i}\left(
V_{i},P_{i},S_{i}\right) \right]  \notag \\
&&-\frac{1-\kappa _{i}}{n}\sum\limits_{j\neq i}\mathbb{E}\left[ \left(
Z_{j}-\tau \left( Z_{j}\right) \right) x_{j}\left(
V_{j},P_{j},S_{j}\right) \right] .
\end{eqnarray}%

When consumers have perfect information, $S_{i}=Z_{i}$, and when they have no information, $S_{i}\equiv \bar{z}$. In both cases the planner, who knows $Z_{i}$ and the average environmental cost $\bar{z}$, de facto knows the consumer's signal, and any function of $\left( Z_{i},S_{i}\right) $ is a function of $Z_{i}$ alone. With intermediate information, the planner does not observe $S_{i}$. It is nevertheless convenient to let the tax schedule depend on both variables, so with a slight abuse of notation in this section we write $\tau \left( Z,S\right) $ instead of $\tau \left( Z\right) $. With intermediate information, this relaxes the planner's problem. If an optimal schedule $\tau \left( Z,S\right) $ does not depend on $S$, it is also optimal among schedules that depend on $Z$ alone. This is the case for aware consumers with homogeneous morality
(Proposition~\ref{thm:existence}). If it does depend on $S$, it cannot be implemented and serves as a benchmark.

We also drop the consumer index $i$ on the i.i.d.\ random variables $Z_{i}$ and $S_{i}$, and accordingly write $\hat{Z}=\mathbb{E}\left[ Z\mid S\right] $ and $\hat{R}=\mathbb{E}\left[ \tau \left( Z,S\right) \mid S\right] $, with realizations $\hat{z}_{s}=\mathbb{E}\left[ Z\mid S=s\right] $ and $\hat{r}_{s}=\mathbb{E}\left[ \tau \left( Z,s\right) \mid S=s\right] =\sum_{z}\Pr \left( Z=z\mid S=s\right) \tau \left( z,s\right) $. Note that $\hat{r}_{s}$ depends on the whole schedule $\tau \left( \cdot ,s\right) $.

As shown in the appendix, the expected average utility at realized outcomes\footnote{Since by assumption $\left( Z,S\right) $ has finite support, $\tau $ can be
represented as a vector in a Euclidean space and $W$ can be viewed as a
real-valued function on that space.}
\begin{equation*}
W\left( \tau \right) =\frac{1}{n}\sum_{i\in I}U_{i}^{\ast }\left(
x_{i}^*,x_{-i}^*\right)
\end{equation*}%
can be written as
\begin{equation}
W\left( \tau \right) =\frac{1}{n}\sum_{i}\mathbb{E}\left[ \int_{A_{i}}^{%
\infty }\left( v-\omega \left( Z\right) -Z-\left( \kappa _{i}-\bar{\kappa}%
\right) \left( Z-\tau \left( Z,S\right) \right) \right) f\left( v\right) dv%
\right] ,  \label{eq:W}
\end{equation}%
where
\begin{equation*}
A_{i}=\omega \left( Z\right) +\tau \left( Z,S\right) +\tilde{\kappa}_{i}\left( \hat{Z}-\alpha \hat{R}\right)
\end{equation*}%
represents the optimal strategy of consumer $i$, namely, to buy if and only
if $V_{i}>A_{i}$. Conditioning on $\left( Z,S\right) $ one may write $%
W\left( \tau \right) $ as a probability weighted sum of terms
\begin{equation*}
W\left( \tau \mid z,s\right) =\frac{1}{n}\sum_{i}\int_{a_{i}\left(
z,s\right) }^{\infty }\left( v-\omega \left( z\right) -z-\left( \kappa _{i}-%
\bar{\kappa}\right) \left( z-\tau \left( z,s\right) \right) \right) f\left(
v\right) dv,
\end{equation*}%
where
\begin{equation*}
a_{i}\left( z,s\right) =\omega \left( z\right) +\tau \left( z,s\right) +\tilde{\kappa}_{i}\left( \hat{z}_{s}-\alpha \hat{r}_{s}\right) .\end{equation*}%
For perfectly informed consumers, $\hat{z}_{s}-\alpha \hat{r}_{s}=s-\alpha \tau \left( s,s\right) $, and for uninformed consumers, whose signal is $s=\bar{z}$, $\hat{z}_{s}-\alpha \hat{r}_{s}=\bar{z}-\alpha \mathbb{E}\left[ \tau \left( Z,\bar{z}\right) \right] $.

Two features of this decomposition will be exploited. First, $W\left( \tau \mid z,s\right) $ depends on the tax function $\tau $ only through its values at the given signal $s$. Hence, the welfare-maximizing tax can be found by maximizing the conditional welfare given $S=s$, $\sum_{z}\Pr \left( Z=z\mid S=s\right) W\left( \tau \mid z,s\right) $, separately for each $s$. Second, if consumers are unaware ($\alpha =0$) or perfectly informed ($S=Z$), then $W\left( \tau \mid z,s\right) $ depends on $\tau $ only through its local value $\tau \left(z,s\right) $, and it suffices to choose $\tau \left( z,s\right) $ to maximize $W\left( \tau \mid z,s\right) $ pointwise for each $\left( z,s\right) $. By contrast, for aware consumers with imperfect information, the values $\tau \left(z,s\right) $ for different $z$ are linked through $\hat{r}_{s}$ in $a_{i}\left( z,s\right)$, and pointwise maximization is not available.

The choice of welfare function in behavioral economics is non-trivial. Since the consumers in our model have a moral concern when making their decisions, we find it natural to let that concern be part of welfare evaluations: both the hedonic pleasure and the fulfillment or violation of the person's own sense of morality are present in the mind of a consumer, and should therefore matter. Each consumer's utility is the material payoff she would obtain in the hypothetical scenario where others, with some probability, use her strategy. Our welfare function evaluates this utility at the realized environmental costs and transfers, rather than at consumers' conditional expectation of them at the time of purchase. The planner thus respects consumers' moral preferences but not their beliefs. Welfare differs from average material payoff at realized outcomes by the term
$-\frac{1}{n}\sum_{i}\left( \kappa _{i}-\bar{\kappa}\right) \mathbb{E}\left[
\left( Z_{i}-\tau \left( Z_{i},S_{i}\right) \right) x_{i}^{\ast }\right] $.
This term vanishes, and welfare equals average material payoff, whenever morality is homogeneous, all consumers use the same strategy, or the tax is Pigovian.
In particular, under homogeneous morality the two criteria coincide for every tax schedule, so the optimal taxes derived in Section~\ref{sec:homogeneity} do not depend on whether moral concern enters welfare.


\subsection{Large population approximation}\label{sec:large population}

The welfare function \eqref{eq:W} is an average over consumers in which consumer $i$ enters only through $\kappa_i$, $\tilde{\kappa}_i$ and the population mean of the degrees of morality. Consider a sequence of populations, indexed by their size $n$, such that the empirical distribution of $(\kappa_1,\ldots, \kappa_n)$ converges weakly to a probability distribution on $[0,1]$ with mean $\bar{\kappa}$. Since $0\le\tilde{\kappa}_i-\kappa_i\le 1/n$, $Z$ has bounded support, and each summand is a bounded and continuous function of $\kappa_i$, $\tilde{\kappa}_i$ and the population mean (recall that $\mathbb{E}[V]<\infty$), per capita welfare converges, for every tax schedule $\tau$, to
\begin{equation}
W\left( \tau \right) =\mathbb{E}\left[ \int_{A}^{\infty }\left( v-\omega
\left( Z\right) -Z-\left( \kappa -\bar{\kappa}\right) \left( Z-\tau \left(
Z,S\right) \right) \right) f\left( v\right) dv\right] ,  \label{eq:Wlim}
\end{equation}%
where $\kappa $ is a random variable with the limit distribution, independent of $\left( Z,S\right) $, the expectation is taken over $\left(
Z,S,\kappa \right) $, and the random threshold for an individual's consumption decision is%
\begin{equation}
A=\omega \left( Z\right) +\tau \left( Z,S\right) +\kappa \left( \hat{Z}-\alpha \hat{R}\right) .\label{V_0}
\end{equation}%
Correspondingly, the conditional welfare given $\left( Z,S\right) =\left(
z,s\right) $ becomes%
\begin{equation*}
W\left( \tau \mid z,s\right) =\mathbb{E}_{\kappa }\left[ \int_{a\left(
z,s,\kappa \right) }^{\infty }\left( v-\omega \left( z\right) -z-\left(
\kappa -\bar{\kappa}\right) \left( z-\tau \left( z,s\right) \right) \right)
f\left( v\right) dv\right] ,
\end{equation*}%
where
\begin{equation}\label{eq:a(z,s,kappa)}
a\left( z,s,\kappa \right) =\omega \left( z\right) +\tau \left( z,s\right) +\kappa \left( \hat{z}_{s}-\alpha \hat{r}_{s}\right) ,
\end{equation}
and the two features of the decomposition noted above continue to hold.

The average per capita consumption is%
\begin{equation}
C\left( \tau \right) =\mathbb{E}\left[ 1-F\left( A\right) \right] ,
\label{Ctz}
\end{equation}%
and the per capita environmental cost of aggregate consumption is%
\begin{equation}
E\left( \tau \right) =\mathbb{E}\left[ Z\left( 1-F\left( A\right) \right) %
\right] .  \label{Etz}
\end{equation}

We will use the large population approximation from this point onward.

\subsection{Moral homogeneity}\label{sec:homogeneity}

When all consumers have the same degree of morality $\kappa <1$, the conditional welfare function is
\begin{equation*}
W\left( \tau \mid z,s\right) =\int_{a\left( z,s\right) }^{\infty }\left(
v-\omega \left( z\right) -z\right) f\left( v\right) dv,
\end{equation*}%
where $a(z,s)=a(z,s,\kappa)$ is given by \eqref{eq:a(z,s,kappa)}.

\begin{proposition}
\label{thm:existence} Assume that consumers have the same degree of morality
$\kappa <1$. Then the welfare maximizing tax exists uniquely and is given by
\begin{equation*}
\tau (z,s)=\left\{
\begin{array}{ll}
z-\kappa\hat{z}_{s} & \text{if }\alpha =0 \\
z & \text{if }\alpha =1.%
\end{array}%
\right.
\end{equation*}
\end{proposition}
Thus, in case consumers are unaware of the tax-transfer system, the optimal tax is equal to the environmental cost reduced by the degree of morality times the consumer's expected environmental cost given their signal.
For example, when consumers have no information regarding the environmental cost
of the good at hand, then $\hat{z}_{s}=\bar{z}$, and the optimal tax becomes
\begin{equation*}
\tau \left( z,s\right) =z-\kappa\bar{z}.
\end{equation*}%
The optimal tax can thus be negative when morality is high and the realized
environmental cost $z$ is smaller than the ex ante expected environmental
cost $\bar{z}$. In effect, the optimal tax then becomes a subsidy that
corrects for the fact that ardent moral consumers, unaware of the
tax-transfer system, overreact to the environmental cost.

When consumers have perfect information about the environmental cost, so that $\hat{z}_{s}=z$, the optimal tax for unaware consumers equals the share $1-\kappa $ of the Pigovian tax: $\tau (z,s)=\left( 1-\kappa \right) z$. This amounts to a ``linear rebate" for morality.
In both boundary cases, the optimal tax for unaware consumers depends on $z$ alone, given the planner's knowledge of which information regime applies.

If consumers are aware of the tax-transfer system, the optimal tax is Pigovian. One might have expected the optimal tax to be lower the more moral consumers are, since such consumers partly internalize the environmental externality. However, tax revenues are returned to the consumer population, and aware moral consumers take this compensating externality into account. Under the Pigovian tax, the purchase of a unit has zero net external effect, so there is no moral dilemma: consumers behave as \emph{homo economicus}, whatever their degree of morality. Moreover, the Pigovian tax does not depend on the signal, so it is optimal also when consumers' information is imperfect and the planner does not know their signals.

Finally, Proposition~\ref{thm:existence} also holds in a finite population, with $\kappa $ replaced by the effective degree of morality $\tilde{\kappa}=1/n+\left( 1-1/n\right) \kappa $.

\subsection{Moral heterogeneity} \label{sec:heterogeneity}

We proceed to analyze optimal taxation when consumer morality is heterogeneous. Consider first the two cases where pointwise maximization is available, that is, unaware consumers ($\alpha =0$) and aware and perfectly informed consumers ($S=Z$). Applying Leibniz' rule to $W\left( \tau \mid z,s\right) $, the necessary first-order condition for an optimal choice of $\tau \left(z,s\right) $ is then

\begin{eqnarray}
0 &=&\mathbb{E}_{\kappa }\left[ \left( \kappa -\bar{\kappa}\right) \left(
1-F\left( a\right) \right) \right]  \label{FOC:heterotax} \\
&&-\mathbb{E}_{\kappa }\left[ \left( 1-\kappa \alpha p_{z|s}\right) \left(
a-\omega \left( z\right) -z-\left( \kappa -\bar{\kappa}\right) \left(
z-\tau \left( z,s\right) \right) \right) f\left( a\right) \right] ,  \notag
\end{eqnarray}%
where $a=a(z,s,\kappa)$ is given in \eqref{eq:a(z,s,kappa)} and $p_{z|s}=\partial \hat{r}_{s}/\partial \tau \left( z,s\right) =\Pr \left( Z=z\mid S=s\right) $, which equals $1$ under perfect information.

Rearranging terms, this implies that an optimal tax must satisfy the following implicit equation:%
\begin{equation}
\tau \left( z,s\right) =z-\left( \hat{z}_{s}-\alpha \hat{r}_{s}\right) H_{1}+H_{2},\label{eq:optimality condition}
\end{equation}%
where
\begin{eqnarray*}
H_{1} &=&\frac{\mathbb{E}_{\kappa }\left[ \kappa \left( 1-\kappa \alpha
p_{z|s}\right) f\left( a\right) \right] }{\mathbb{E}_{\kappa }\left[ \left(
1+\kappa -\bar{\kappa}\right) \left( 1-\kappa \alpha p_{z|s}\right) f\left(
a\right) \right] } \\
H_{2} &=&\frac{\mathbb{E}_{\kappa }\left[ \left( \kappa -\bar{\kappa}\right)
\left( 1-F\left( a\right) \right) \right] }{\mathbb{E}_{\kappa }\left[
\left( 1+\kappa -\bar{\kappa}\right) \left( 1-\kappa \alpha p_{z|s}\right)
f\left( a\right) \right] }.
\end{eqnarray*}


To interpret the optimality condition under moral heterogeneity, consider unaware consumers ($\alpha =0$). Let $\kappa _{m}=\mathbb{E}_{\kappa }\left[
\kappa f\left( a\right) \right] /\mathbb{E}_{\kappa }\left[ f\left( a\right) \right] $ denote the average degree of morality among marginal consumers, those whose valuation equals their threshold. Then $H_{1}=\kappa_{m}/\left( 1+\kappa _{m}-\bar{\kappa}\right) $. This term lowers the tax because marginal consumers already internalize part of the environmental cost. The term $H_{2}$ arises only under heterogeneity. Its numerator is the covariance between a consumer's morality and her probability of buying. It captures a direct effect of the tax on welfare: by \eqref{eq:Wlim}, a marginal tax increase raises the price by one unit and reduces its net external cost $Z-\tau $ by one unit, to which welfare attaches the weight $1+\kappa -\bar{%
\kappa}$, so the direct effect per purchase is $\kappa -\bar{\kappa}$. Since
more moral consumers buy less, the covariance is negative,\footnote{%
For $\alpha =0$, the threshold $a\left( z,s,\kappa \right) $ is strictly
increasing in $\kappa $ whenever $\hat{z}_{s}>0$, so, when thresholds are
positive, $1-F\left( a\left( z,s,\kappa \right) \right) $ is strictly
decreasing in $\kappa $. Its covariance with $\kappa $ is therefore strictly
negative when morality is heterogeneous. The common denominator of $H_{1}$
and $H_{2}$ equals $\left( 1+\kappa _{m}-\bar{\kappa}\right) \mathbb{E}_{\kappa }[ f( a) ]$, which is positive.} and so is $H_{2}$.
Both terms thus pull in the same direction: for unaware consumers, any optimal tax lies below the Pigovian tax at every $\left( z,s\right) $ with $\hat{z}_{s}>0$. If $\hat{z}_{s}=0$, the first-order condition gives the Pigovian tax.

When consumers are aware, the pointwise first-order condition \eqref{FOC:heterotax} is exact only under perfect information. For general signals, the values $\tau \left( z,s\right) $ for different $z$ are linked through $\hat{r}_{s}$, and the first-order condition acquires additional terms; it is derived in the Appendix. Nevertheless, the Pigovian tax satisfies the generalized first-order condition in all cases.

\begin{theorem}
\label{P existence}Suppose consumers are aware of the tax-transfer system ($%
\alpha =1$). Then Pigovian taxation, $\tau ^{\ast }\left( z\right) =z$ for
all $z$, satisfies the necessary first-order condition for welfare
maximization, irrespective of the distribution of morality and of
consumers' information.
\end{theorem}
Theorem~\ref{P existence} also holds in a finite population.
\subsection{Optimality of the Pigovian tax}

We now establish, under additional regularity conditions, that the Pigovian
tax is not only a solution to the first-order condition (Theorem~\ref{P
existence}) but is in fact the unique optimum.

\begin{theorem}\label{thm:unique_tax_aware_and_informed}
Suppose consumers are aware of the tax-transfer system ($\alpha =1$) and have perfect information ($S=Z$). Assume the hedonic valuation distribution is exponential, $F\left( v\right) =1-e^{-v/\mu }$ for some $\mu >0$, and that the support of $\kappa $ is contained in $\left[ 0,\kappa_{max}\right] $ for some $\kappa_{max}<2\sqrt{2}-2$. Then the Pigovian tax $\tau \left( z\right) =z$ is the unique optimal tax among tax schedules under which consumer prices are nonnegative, $\omega \left( z\right) +\tau \left( z\right) \geq 0$ for all $z$.
\end{theorem}

The neutrality result for aware consumers in Proposition~\ref{thm:existence} depends crucially on the assumption that tax revenues are returned lump-sum without friction. Suppose that only a fraction $\lambda \in (0,1)$ is returned. The consumer's threshold is then obtained by replacing $\alpha$ with $\alpha\lambda$ in \eqref{V_0}, while the planner must also account for the loss $(1-\lambda)\tau$ per unit sold. With homogeneous morality, exponential valuations and perfect information, the optimal tax is
\begin{equation*}
\tau(z) = \frac{z}{\lambda}-\frac{(1-\lambda)\mu}{\lambda(1-\kappa)(1-\kappa\lambda)}
\quad\text{and}\quad
\tau(z) = \frac{(1-\kappa)z-(1-\lambda)\mu}{\lambda}
\end{equation*}
for aware and unaware consumers, respectively. Any waste of revenues thus restores a role for morality, even for aware consumers. Two effects are at work. First, the lost revenue $(1-\lambda)t$ on a purchase taxed at $t$ is itself an external cost, so a tax that internalizes the full marginal social cost satisfies $t=z+(1-\lambda)t$, that is, $t=z/\lambda$. Second, a fraction $1-\lambda$ of the tax paid on inframarginal units is lost as well, and lowering the tax reduces this loss at the cost of a small distortion at the margin. This pulls the tax below $z/\lambda$ by an amount proportional to $\mu$, the expected surplus of an inframarginal buyer. The first effect dominates when $z$ is large relative to $\mu$: the optimal tax exceeds its value without waste ($z$ for aware and $(1-\kappa)z$ for unaware consumers) if and only if $z>\mu/\left((1-\kappa)(1-\kappa\lambda)\right)$ and $z>\mu/(1-\kappa)$, respectively. For unaware consumers, the first effect operates only on the part $(1-\kappa)z$ of the environmental cost that they do not already internalize.

The upper bound on morality  $\kappa_{max}<2\sqrt{2}-2 \approx 0.83$ may not be binding in practice. Laboratory studies \citep{van_leeuwen_estimating_2024} estimate the distribution of morality to have a mean of $0.13$, a standard deviation of $0.14$, and an upper bound of $0.8$.

We finally note that under a material welfare criterion  \citep[used in][]{daube_moral_2016}, Pigovian taxation of aware consumers is globally optimal for any valuation distribution, morality distribution, and signal structure, because it implements the first best. The additional conditions in Theorem \ref{thm:unique_tax_aware_and_informed} are needed because the planner maximizes welfare including consumers'  moral-utility term.

\section{Comparative statics \label{sec:compstat}}

We proceed to present some comparative statics results. In this part of the
study, we will refer to the following additional assumption:

\begin{quote}
\textbf{[M]}: \label{A:density}The probability density function $f$ of the
valuation distribution is strictly decreasing at valuations above the
minimal production cost.
\end{quote}

This assumption holds for any strictly unimodal distribution on $\mathbb{R}_{+}$ with
positive density with mode at or below the minimal production cost. It holds
trivially when the value distribution is exponential. Under the assumption, $%
1-F\left( v\right) $ is a strictly convex function of $v$ for $v\geq
\min_{z\in \mathcal{Z}}\omega \left( z\right) $, where $\mathcal{Z}$ is the
finite support of $Z$.

\subsection{Moral heterogeneity}

We proceed to explore the effect of moral heterogeneity in the consumer
population. We will compare the case where the distribution of $\kappa$ is
non-degenerate with mean value $\bar{\kappa}$ with the case where all
consumers have the same morality $\bar{\kappa}$.

Consumption in a heterogeneous population under any taxation $\tau \geq 0$
and environmental cost $z\geq 0$ is $C\left( \tau \right) $, as defined in %
\eqref{Ctz}. Consumption under homogeneous morality, $C_{\hom }\left( \tau
\right) $, is defined in the same way, except in that case the threshold valuation in \eqref{V_0} reduces to
\begin{equation*}
A=\omega \left( Z\right) +\tau \left( Z,S\right) +\bar{\kappa}\left( \hat{Z}-\alpha \hat{R}\right) .
\end{equation*}%
By Jensen's inequality and the convexity of $1-F$ implied by assumption [M],
heterogeneity with respect to morality leads to both increased consumption
and increased environmental cost of consumption:

\begin{proposition}
\label{prop:consumption_heteromoral}Let assumption [M] hold and let the
distribution of morality be non-degenerate. Consider a tax scheme $\tau
\left( Z,S\right) \geq 0$ such that $\hat{Z}-\alpha \hat{R}$ is nonnegative almost surely and positive with positive probability. Then%
\begin{equation*}
C_{\hom }\left( \tau \right) <C\left( \tau \right) \text{ and }E_{\hom
}\left( \tau \right) <E\left( \tau \right) .
\end{equation*}
\end{proposition}

The condition on $\hat{Z}-\alpha \hat{R}$ holds for unaware consumers under any nonnegative tax, and for aware consumers whenever the expected tax does not exceed the expected
environmental cost. It rules out, as a boundary case, aware consumers facing the Pigovian tax. For such consumers morality does not affect behavior, and the inequalities are replaced by equalities.

\subsection{Information}
The purpose of environmental labels on consumer goods is usually to reduce environmental harm by reducing the consumption of the most harmful goods. We therefore ask how consumption and its environmental cost respond when consumers are given free information about the environmental cost of individual goods.

We consider consumers with no, intermediate, and perfect information under an arbitrary tax scheme $\tau $ that depends on $Z$ (but not on $S$) and write $\bar{\tau}=\mathbb{E}\left[ \tau \left( Z\right) \right] $. From \eqref{V_0}, the random consumption threshold in the three information regimes is
\begin{equation*}
A\left( \tau ,S\right) =\left\{
\begin{array}{ll}
P+\kappa \left( \bar{z}-\alpha \bar{\tau}\right) & \text{if }S\equiv\bar z \\
P+\kappa \left( \hat{Z}-\alpha \hat{R}\right) & \text{general }S \\
P+\kappa \left( Z-\alpha \tau \left( Z\right) \right) & \text{if }S=Z.%
\end{array}%
\right.
\end{equation*}

By \eqref{Ctz} and \eqref{Etz}, the expected consumption and the expected
environmental cost of consumption are, respectively,
\begin{equation*}
C(\tau,S)=\mathbb{E }[1-F(A(\tau,S))] \text{ and } E(\tau,S) = \mathbb{E }%
[Z(1-F(A(\tau,S)))].
\end{equation*}

First, suppose that consumers are aware of the
tax-transfer system and the tax is the Pigovian tax. Then $C\left( \tau
,S\right) =\mathbb{E}\left[ 1-F\left( P\right) \right] $ and $E\left( \tau
,S\right) =\mathbb{E}\left[ Z\left( 1-F\left( P\right) \right) \right] $, so aggregate consumption and its environmental cost do not depend on information, and environmental labeling serves no purpose. For unaware consumers with a common degree of morality, the same holds if the planner re-optimizes the tax when the information regime changes, since the optimal tax $z-\kappa \hat{z}_{s}$ depends on the regime (Proposition \ref{thm:existence}). This is possible under perfect or no information; under intermediate information, the optimal tax for unaware consumers depends on the signal $s$ and cannot be implemented.

Second, consider the case of constant production cost and no tax. In this case, information affects consumption only through the moral term in the threshold, and convexity of the demand function implies (by Jensen's inequality) that consumption is lowest under no information and highest under perfect information.

\begin{proposition}
\label{prop:info}Suppose that assumption [M] holds, that there is no tax, and that the production cost does not depend on the environmental cost. Then%
\begin{equation*}
C\left( 0,\bar{z}\right) \leq C\left( 0,S\right) \leq C\left( 0,Z\right) ,
\end{equation*}%
and $C\left( 0,\bar{z}\right) <C\left( 0,Z\right) $ if $Z$ is non-degenerate and the average degree of morality is positive.
\end{proposition}

The following example illustrates how increased environmental  information to unaware consumers may cause increased or decreased environmental harm, depending on the range of environmental costs.

\begin{example}
Suppose that all consumers are unaware ($\alpha =0$) and have the same positive degree of morality $\kappa>0$. Let the consumer price $p$ be independent of environmental cost $Z$ and suppose that there is no tax. If the valuation distribution is exponential with mean $\mu$, then the aggregate environmental cost  $E\left( 0 ,Z\right) =\mathbb{E}\left[ g\left( Z\right) \right] $ and $E\left( 0 ,\bar{z}\right) =g\left( \bar{z}\right) $, where
\begin{equation*}
g\left( z\right) =z\left( 1-F\left( p+\kappa z\right) \right) =ze^{-\left( p+\kappa z\right) /\mu }.
\end{equation*}%
Since 
$$g^{\prime \prime }\left( z\right) =\frac{\kappa }{\mu }e^{-\left( p+\kappa z\right) /\mu }\left( \frac{\kappa z}{\mu }-2\right), $$ 
the function $g$ is strictly concave on $\left[ 0,2\mu /\kappa \right] $ and strictly convex on $\left[ 2\mu /\kappa ,\infty \right) $. By Jensen's inequality, if $Z$ is non-degenerate with support in $\left[ 0,2\mu /\kappa \right] $, then information strictly decreases the expected environmental cost of consumption, and if its support is in $\left[ 2\mu /\kappa ,\infty \right) $, then information strictly increases it. Since $2\mu /\kappa \geq 2\mu $, information reduces the expected environmental cost whenever environmental costs are below twice the mean hedonic valuation.
\end{example}

\subsection{Awareness}\label{sec:awareness}

Awareness of the tax system reduces the consumption threshold of moral consumers and hence increases aggregate consumption, ceteris paribus. On the other hand, awareness tends to increase the optimal tax. One may then ask about the combined effect: does awareness increase or decrease consumption under the associated optimal tax? It turns out that awareness has no effect on consumption, provided consumers have the same degree of morality.

\begin{proposition}
\label{prop:awareness and morality} Suppose that consumers all have the same degree of morality and that the tax is the optimal tax. Then the consumption of the dirty good is independent of whether consumers are aware of the tax-transfer system or not.
\end{proposition}

\section{A quota system\label{sec: quota system}}

The European emissions trading system has delivered substantial reductions in GHG emissions \citep{bayer_european_2020}. Can our model shed light on how moral consumers behave under a quota system? Suppose, thus, that a fixed number of allowances $q>0$ is issued per capita, and that an allowance is required for each unit of environmental cost $Z$. Suppose that the quota system covers the varieties of goods considered in our model. Under perfect competition,  producers' payment for allowances acts like a tax $\tau \left(Z\right) =\gamma Z$ on each unit consumed, $P=\omega(Z) + \gamma Z$, where $\gamma $ is the allowance price paid by producers, and the allowance revenue is returned to consumers. Let $E\left( \gamma \right) $ denote expected per capita emissions at allowance price $\gamma $; for consumers unaware of the quota system, this is \eqref{Etz} evaluated at $\tau \left( Z\right) =\gamma Z$. As we shall see, the optimal strategy, and hence $E\left( \gamma \right) $, depends on whether consumers are aware of the quota system, so we specify $E\left( \gamma \right) $ separately in the two cases below. The quota requires $E\left( \gamma \right) \leq q$, and the allowance price adjusts to clear the market for allowances. We assume throughout that the quota binds, so that $E\left( \gamma \right) =q$ at a positive allowance price $\gamma >0$. 

\subsection{Consumers unaware of the quota system}

If all consumers are unaware of the quota system, then, effectively, they
are in the same situation as under a tax. Thus, their optimal strategy is
still $x^{\ast }$ \eqref{x*}. The allowance price $\gamma $ adjusts
to satisfy the constraint
\begin{equation*}
E(\gamma )=\mathbb{E}\left[ Z\left( 1-F\left( \omega \left( Z\right) +\gamma
Z+\kappa \hat{Z}\right) \right) \right] \leq q.
\end{equation*}%
Granted that the constraint is binding, it is easily verified that the allowance price falls when the cap is raised and rises when it is reduced:
\begin{equation*}
\frac{\partial \gamma }{\partial q}=-\frac{1}{\mathbb{E}\left[ Z^{2}f\left(
\omega \left( Z\right) +\gamma Z+\kappa \hat{Z}\right) \right] }<0.
\end{equation*}%
Thus, the welfare optimal quota is the one that implies the welfare optimal
allowance price. Finding the welfare optimal allowance price is a slightly
different problem than finding the welfare optimal tax $\tau $. We can, however, still obtain conclusions
regarding the optimal quota and allowance price by applying results already established in this paper. Proposition \ref{thm:existence} provided the optimal tax for a perfectly informed, morally homogeneous consumer population. This leads immediately to

\begin{proposition}
\label{prop:quota_unaware}Consider a consumer population that is unaware of the quota system but perfectly informed about the environmental costs of goods, and has a common degree of morality $\bar{\kappa}$. The optimal quota corresponds to the allowance price $\gamma = 1-\bar{\kappa}$.
\end{proposition}

Proposition \ref{prop:consumption_heteromoral} established that, for unaware consumers and any nonnegative tax, a morally homogeneous consumer population causes a strictly lower environmental cost than a morally heterogeneous one.
Hence,

\begin{proposition}
Let assumption [M] hold and let the distribution of morality be non-degenerate. For consumers unaware of the quota system, at any binding quota $q$, the allowance price is higher for a morally heterogeneous consumer population than for a morally homogeneous consumer population.
\end{proposition}

\subsection{Consumers aware of the quota system}\label{sec:quota:aware}

We finally turn to the case where consumers are aware of the quota system. Such consumers understand that the allowance price adjusts to clear the market for allowances and that the allowance revenue is returned uniformly to the consumer population. To choose her strategy, a consumer evaluates each alternative strategy in a hypothetical scenario where a share $\kappa_i$ of the other consumers also adopts it. She must therefore determine the allowance price that would clear the market in that hypothetical situation, and the resulting prices and emissions. A change in the allowance price changes the price of each unit in proportion to its environmental cost. The consumer therefore needs to know the distribution $F$ of valuations, the joint distribution of environmental costs and signals $\left(Z,S\right)$, and the production cost function $\omega$. This departs from our maintained assumption in Section~\ref{sec:information}, under which consumers do not know $\omega$ and regard the price as uninformative about the environmental cost. In the rest of this section, all expectations are taken with respect to the true joint distribution of $\left(V,Z,S\right)$.
Under perfect information this makes no difference, since the signal then reveals the environmental cost and both assumptions lead to the same conditional expectations. Moreover, the equilibrium strategy in Proposition~\ref{prop:quota_aware} below, to buy if and only if the valuation exceeds the price, does not involve the consumer's expectation of the environmental cost. The modified assumption matters only for how consumers evaluate alternatives to that strategy.

Let%
\begin{equation*}
Q\left( \gamma \right) =\mathbb{E}\left[ Z\left( 1-F\left( \omega \left(
Z\right) +\gamma Z\right) \right) \right]
\end{equation*}%
denote per capita emissions at allowance price $\gamma $ if every consumer behaved like \emph{homo economicus}, that is, bought a unit if and only if her valuation exceeds the price. The Pigovian tax corresponds to an allowance price of $1$; call the resulting level of emissions, $q^{\ast }=Q\left( 1\right) $, the \emph{Pigovian quota}.

\begin{proposition}
\label{prop:quota_aware}Consider consumers who are aware of the quota system, with arbitrary signals and degrees of morality below unity, and let the quota satisfy $q\leq q^{\ast }$. Then there exists an equilibrium in which every consumer buys if and only if $v>p$. The allowance price $\gamma \geq 1$ is then the unique solution to
\begin{equation}
q=Q\left( \gamma \right) .  \label{tau q}
\end{equation}
This equation is independent of consumers' morality and signals. In this equilibrium, the strategy of each consumer is almost surely her unique best response.
\end{proposition}

The condition $q\leq q^{\ast }$ cannot be dropped. If
$q>q^{\ast }$, the market-clearing price under the strategy in Proposition
\ref{prop:quota_aware} is below $1$, and a sufficiently moral consumer then prefers a strategy under which, were others to follow it, the quota would not bind. The proof in the Appendix establishes this for perfectly informed consumers whose degree of morality is sufficiently close to one.

\begin{theorem}
\label{prop:quota_optimal}Consider consumers who are aware of the quota system, with arbitrary signals and degrees of morality below unity.
Among quotas $q\leq q^{\ast }$, the welfare-optimal quota is $q^{\ast }$,
with equilibrium allowance price $\gamma =1$. The optimal quota thus
reproduces the allocation under Pigovian taxation.
\end{theorem}

\section{Extensions}\label{sec:extensions}
\subsection{Continuous demand}\label{rem:continuous}

The unit-demand assumption is less restrictive than it may seem. Suppose instead that each consumer chooses a quantity $x_{i}\in \left[ 0,1\right] $ of the good and that her hedonic utility is $u\left(
x_{i}\right) -p_{i}x_{i}$ plus the transfer and environmental terms in Section \ref{sec:hedonic}, where%
\begin{equation*}
u\left( x\right) =\int_{0}^{x}F^{-1}\left( 1-s\right) ds.
\end{equation*}%
Then $u\left( 0\right) =0$, $u\left( 1\right) =\mathbb{E}\left[ V\right] $, and $u^{\prime }$ decreases continuously from $\infty $ to $0$, so the demand function $D=\left( u^{\prime }\right) ^{-1}$ equals $1-F$ on $\mathbb{R}_{+}$ and $1$ on $\mathbb{R}_{-}$. Conversely, $F$ can be recovered from $u$ as $F=1-(u')^{-1}$ on $\mathbb R_+$.

In a finite population, consumer $i$'s moral utility is then $$u\left( x_{i}\right) -\left( P_{i}+\tilde{\kappa}_{i}( \hat{Z}_{i}-\alpha \hat{R}_{i}) \right) x_{i}$$ plus terms she does not influence, where $\tilde{\kappa}_{i}=1/n+\left( n-1\right) \kappa _{i}/n$ as before. Hence, instead of her optimal binary strategy $x^*$ in Lemma \ref{opt x}, she demands $x_i = D\left( A_{i}\right) $, where $A_{i}=P_{i}+\tilde{\kappa}_{i}( \hat{Z}_{i}-\alpha \hat{R}_{i}) $ is the threshold for $V_{i}$ implied by \eqref{x*}, and Proposition \ref{prop:equilibrium} holds as stated. Substituting $s=1-F\left( v\right) $ in the definition of $u$ shows that, for all $a\in \mathbb{R}$ and $c\in \mathbb{R}$,
\begin{equation*}
u\left( D\left( a\right) \right) -cD\left( a\right) =\int_{a}^{\infty
}\left( v-c\right) f\left( v\right) dv,
\end{equation*}%
so her hedonic utility net of any per-unit cost $c$ equals the corresponding expectation over $V$ under unit demand, and her quantity $D\left( A_{i}\right) $ equals the probability of purchase $1-F\left(A_{i}\right) $.

Consequently, the welfare, consumption and environmental cost expressions derived in Section \ref{sec:optimal taxation}, and all results in Sections \ref{sec:optimal taxation} and \ref{sec:compstat}, hold verbatim for continuous demand. The results in Section \ref{sec:quota:aware} also extend, with strategies
taking values in $\left[ 0,1\right] $; we omit the details. Exponential valuations correspond to $u\left(x\right) =\mu x\left( 1-\ln x\right) $, and assumption [M] in Section \ref{sec:compstat} to convexity of the demand function on $\mathbb{R}_{+}$.

\subsection{Convex environmental cost}\label{sec:ext:convex}

Another restriction in our unit demand model is the assumption that the environmental cost is linear in consumption. In order to relax this restriction, we now distinguish between the per capita \emph{environmental load} $\ell=\frac{1}{n}\sum_{i\in I}b_{i}Z_{i}$, the quantity in \eqref{Etz}, and the \emph{environmental cost} $\psi\left( \ell\right) $ that it imposes on each individual, where $\psi:\mathbb{R}_{+}\rightarrow \mathbb{R}_{+}$ is continuously differentiable, increasing and strictly convex with $\psi\left( 0\right) =0$. In the linear case $\psi\left( \ell\right) =\ell$ the two coincide, and this is the model of the preceding sections; here each individual $j\in I$ experiences the disutility $\psi\left( \ell\right) $ in place of $\ell$ in \eqref{ui}.  The tax-transfer scheme is unchanged.

Two things change. First, the game is no longer additively separable across consumers, so a consumer's optimal strategy depends on the strategies of others, and Lemma \ref{opt x} no longer applies as stated. Second, the marginal environmental cost $\psi^{\prime }\left( \ell\right) $ is no longer constant. Since a \emph{homo moralis} consumer $i$ evaluates her strategy at the counterfactual load that would arise if a share $\kappa _{i}$ of the other consumers were to use it, the marginal cost she perceives depends on her own strategy and on the strategies of others, and it differs in general from the marginal cost at the actual load.

We show that the threshold structure of optimal strategies nevertheless survives, that equilibrium still exists and is unique in a large population, and that the neutrality result for aware consumers extends.

To obtain these results, we work directly with the large population limit of Section \ref{sec:large population}. For a strategy $x\in X$, let
\begin{equation*}
\ell\left( x\right) =\mathbb{E}\left[ x\left( V,P,S\right) Z\right] \quad \text{and}\quad r\left( x\right) =\mathbb{E}\left[ x\left( V,P,S\right) \tau \left( Z\right) \right]
\end{equation*}
denote the per capita environmental load and the per capita tax revenue that would result if all consumers used strategy $x$. Let $\bar{\ell}$ and $\bar{r}$ denote the actual per capita load and revenue in the population. By the law of large numbers applied to the Bernoulli variables $K_{ij}$ in \eqref{ES U}, the moral utility of consumer $i$ with strategy $x_{i}$ converges, as $n\rightarrow \infty $, to
\begin{align}
\tilde{U}_{i}\left( x_{i};\bar{\ell},\bar{r}\right) &=\mathbb{E}\left[ \left( V-P\right) x_{i}\left( V,P,S\right) \right]  \notag\\
&-\psi\left( \kappa _{i}\ell\left( x_{i}\right)+\left( 1-\kappa _{i}\right) \bar{\ell}\right) +\alpha \left( \kappa _{i}r\left( x_{i}\right) +\left( 1-\kappa _{i}\right) \bar{r}\right) .\label{eq:U convex}
\end{align}
Her moral utility depends on the strategies of others only through the aggregates $\bar{\ell}$ and $\bar{r}$. Since $\bar{r}$ enters additively, only $\bar{\ell}$ affects her choice of strategy.
The effect of her own purchase on the load vanishes at rate $1/n$, as in Lemma \ref{opt x}, but the counterfactual load $\tilde{\ell}_{i}=\kappa _{i}\ell\left( x_{i}\right) +\left( 1-\kappa _{i}\right) \bar{\ell}$ does not.

For $m\geq 0$, let $x_{i}^{(m)}$ denote the strategy that buys if and only if $V_{i}-P_{i}>\kappa _{i}\left( m\hat{Z}_{i}-\alpha \hat{R}_{i}\right) $, and let $\ell_{i}^{(m)}=\ell\left( x_{i}^{(m)}\right) $ be the load it would generate if all consumers used it.

To characterize optimal strategies, we allow strategies to take values in $\left[ 0,1\right] $ rather than $\left\{ 0,1\right\} $. The utility \eqref{eq:U convex} is concave on this convex set of strategies, so optimality can be characterized by a first-order condition. The relaxation is without loss of generality: by the following lemma, optimal strategies take values in $\left\{ 0,1\right\} $ almost surely.

\begin{lemma}
\label{lem:convex_opt}Let $\bar{\ell}$ and $\bar{r}$ be given. Consumer $i$ has an optimal strategy among strategies taking values in $\left[ 0,1\right] $, and it is almost surely unique. It is, almost surely,
\begin{equation}\label{eq:x convex}
x_{i}^{\ast }\left( V_{i},P_{i},S_{i}\right) =\left\{
\begin{array}{ll}
1 & \text{if }V_{i}-P_{i}>\kappa _{i}\left( m_{i}\hat{Z}_{i}-\alpha \hat{R}_{i}\right) \\
0 & \text{otherwise,}
\end{array}
\right.
\end{equation}
where $m_{i}\geq 0$ is the unique solution to $m_{i}=\psi^{\prime }\left( \kappa _{i}\ell_{i}^{(m_{i})} +\left( 1-\kappa _{i}\right) \bar{\ell}\right) $.
\end{lemma}

Thus optimal strategies retain the threshold form of Lemma \ref{opt x}, with the expected environmental cost $\hat{Z}_{i}$ multiplied by $m_{i}=\psi^{\prime }\left( \tilde{\ell}_{i}\right) $. This is the marginal damage at the counterfactual load $\tilde{\ell}_{i}=\kappa _{i}\ell\left( x_{i}^{\ast }\right) +\left( 1-\kappa _{i}\right) \bar{\ell}$ that would arise if a share $\kappa _{i}$ of the population used her strategy. The nonlinearity of $\psi$ enters the consumer's problem only through this scalar.

By Lemma \ref{lem:convex_opt}, consumer $i$'s best response depends on the strategies of others only through $\bar{\ell}$, and it is monotone: a higher $\bar{\ell}$ raises $m_{i}$ and lowers her own load $\ell\left( x_{i}^{\ast }\right) $.
Consequently, the per capita load generated by all consumers' best responses is a continuous and decreasing function of $\bar{\ell}$ on $\left[ 0,\mathbb{E}\left[ Z\right] \right] $, and it therefore has exactly one fixed point. In sum, in the large population the game has a unique Nash equilibrium, in which every consumer uses a threshold strategy of the form \eqref{eq:x convex}.

Consider next optimal taxation of aware consumers ($\alpha =1$). The first-best allocation maximizes $\mathbb{E}\left[ \left( V-\omega \left( Z\right) \right) x\right] -\psi\left( \ell\left( x\right) \right) $ over allocations $x$ that depend on $\left( V,Z\right) $ and take values in $\left[ 0,1\right] $. By the argument in the proof of Lemma~\ref{lem:convex_opt}, it is almost surely unique and given by
\begin{equation*}
x^{\ast }=\mathbf{1}\left\{ V>\omega \left( Z\right) +\psi^{\prime }\left( \ell^{\ast }\right) Z\right\} ,
\end{equation*}
where the first-best load $\ell^{\ast }$ is the unique solution to
\begin{equation*}
\ell^{\ast }=\mathbb{E}\left[ Z\,\mathbf{1}\left\{ V>\omega \left( Z\right) +\psi^{\prime }\left( \ell^{\ast }\right) Z\right\} \right] .
\end{equation*}

\begin{proposition}
\label{prop:convex_pigou}Suppose that consumers are aware of the tax-transfer system ($\alpha =1$), and let the tax be Pigovian at the first-best marginal damage, $\tau \left( z\right) =\psi^{\prime }\left( \ell^{\ast }\right) z$. Then, irrespective of the distribution of morality and of consumers' information, the unique equilibrium is that every consumer buys if and only if $V_{i}>\omega \left( Z_{i}\right) +\psi^{\prime }\left( \ell^{\ast }\right) Z_{i}$. The induced allocation is the first best, and at this allocation each consumer's moral utility equals her material utility.
\end{proposition}

Thus the neutrality result of Section \ref{sec:optimal taxation} extends to convex environmental cost: for aware consumers, whatever their information, the Pigovian tax evaluated at the first-best marginal damage neutralizes morality and implements the first-best allocation.

\subsection{Mixed awareness}\label{sec:ext:awareness}

Proposition \ref{prop:awareness and morality}, like the rest of the paper, presumes that either all consumers are aware of the tax-transfer system or all are unaware. We now ask what happens when awareness is mixed. Suppose that a share $\phi \in \left( 0,1\right) $ of consumers is aware of the tax-transfer system and the rest are not, with awareness independent of valuations, environmental costs, signals and morality. Awareness affects behavior but not utility at realized outcomes, which includes the actual transfer whether or not consumers are aware of it when making their buying decisions, so the welfare function is as before, with the threshold \eqref{V_0} evaluated at $\alpha =1$ for the aware and at $\alpha =0$ for the unaware consumers.

\begin{proposition}
\label{prop:mixed_awareness}Suppose that consumers have the same degree of
morality $\kappa \in (0,1)$, that they are perfectly informed ($S=Z$), and that a
share $\phi \in \left( 0,1\right) $ of them is aware of the tax-transfer
system. An optimal tax schedule $\tau$ satisfies%
\begin{equation*}
\left( 1-\kappa \right) z<\tau \left( z\right) <z, \quad \text{for all } z>0.
\end{equation*}%
Under an optimal tax, aware consumers face a lower and unaware consumers a
higher threshold than $\omega \left( z\right) +z$. Welfare is strictly lower than in a population where all consumers are aware or all are unaware.
\end{proposition}

By contrast, with uniform awareness, a single tax aligns every consumer's threshold with the first-best threshold $\omega \left( z\right) +z$, whether that tax is the Pigovian tax or the morality-adjusted tax $\left( 1-\kappa \right) z$. With mixed awareness, the two groups respond differently to the same tax, and no single tax can align both. The planner then compromises: the optimal tax lies between the two pure-case optima, aware consumers consume too much and unaware consumers too little, and heterogeneity in awareness is costly in welfare terms even when morality is homogeneous.

\subsection{Other externalities}\label{sec:ext:other}

We have framed the analysis in terms of an environmental harm with the following features. First, a purchase imposes a cost that is borne by all $n$ individuals in the population, in equal shares. Second, the cost is additive across units, a feature relaxed in Section \ref{sec:ext:convex}. Third, the cost is attached to a characteristic $Z_{i}$ of the unit at hand which the social planner observes and the buyer may know only through a signal. Fourth, the cost is a hedonic cost to the consumer herself: she bears her share of the harm from every unit bought, including the units that others would buy if they used her strategy. This is essential, since a \emph{homo moralis} consumer evaluates her strategy by the material payoff she would obtain if others were to use it; an externality that she does not experience herself does not enter her moral calculus.

Any externality with these features is covered by our analysis without change. Besides greenhouse gas emissions, examples include waste and litter, noise, and public-health effects of consumption. Harms that do not fall on members of the population, such as harm to animals or to future generations, are covered to the extent that the consumers themselves experience disutility from them; $Z_{i}$ is then the disutility that a purchase causes to each member of the population, whether directly or through concern for third parties.

Positive externalities correspond to $Z_{i}<0$  (we thus relax the so far maintained assumption that $Z$ is always nonnegative). Examples include vaccination, the adoption of technologies with network or learning spillovers, and the consumption of goods whose production yields public benefits. The equilibrium analysis of Section \ref{sec:model} and the results in Sections \ref{sec:large population}--\ref{sec:heterogeneity} do not use the sign of $Z$ and carry over with the sign reversed, with one qualification: uniqueness in Proposition \ref{thm:existence} requires $\omega \left( z\right) +z>0$, that is, a positive social cost of the unit. When the external benefit exceeds the production cost, it is optimal that every consumer buys, and any subsidy large enough to bring this about is optimal; the tax schedules in Proposition \ref{thm:existence} remain among the optima.
For aware consumers, the Pigovian subsidy $\tau \left( z\right) =z<0$ satisfies the first-order condition for optimality irrespective of the distribution of morality (Theorem~\ref{P existence}), and it is optimal under moral homogeneity (Proposition~\ref{thm:existence}), uniquely so when $\omega \left( z\right) +z>0$.
For unaware, perfectly informed consumers with homogeneous morality, the optimal subsidy is $\tau \left( z\right) =\left( 1-\kappa \right) z$, which is smaller in absolute value than the Pigovian subsidy: moral consumers already buy more of the good, and the subsidy needed to correct the remaining externality is correspondingly smaller. Under moral heterogeneity both terms $H_{1}$ and $H_{2}$ in \eqref{eq:optimality condition} work in the same direction, since more moral consumers now buy more, so the covariance in $H_{2}$ is positive. The comparative statics of Section \ref{sec:compstat} and the quota analysis of Section \ref{sec: quota system} rely on $Z\geq 0$ and would require separate treatment.

Externalities that fall on buyers only, such as congestion, do not fit the first feature. When a moral consumer contemplates others adopting her strategy, the counterfactual then changes not only who buys but also who is harmed, and the aggregative structure of the game in Section \ref{sec:hedonic} is lost. We leave such cases for future work.

\section{Related literature}\label{sec:related literature}

Our paper relates to three strands of literature: models of Kantian morality applied to environmental policy, studies of how moral motivation interacts with the choice of policy instrument, and other models of moral motivation, including some that explain why incentives sometimes backfire.

Closest to our paper are \citet{daube_moral_2016} and \citet{eichner_kantians_2022}. In both papers, a consumer's moral preferences combine her selfish objective with a Kantian one, in which everybody chooses her quantity. In \citet{daube_moral_2016}, the Kantian objective counts the returned tax revenue, so the Pigovian tax induces efficient consumption whatever the distribution of morality; they note that this fails if consumers ignore the government budget constraint. \citet{eichner_kantians_2022} study that case and show that the taxes that are efficient for material welfare are then type-specific, below the conventional Pigovian tax and lower for more moral consumers. We instead derive consumer behavior from equilibrium strategies in a game of incomplete information among a finite population of \emph{homo moralis} consumers. A consumer evaluates her strategy, a rule that specifies whether to buy given her valuation, the price and her signal, by the material payoff she would obtain if each other consumer, with probability $\kappa$, applied it to their own valuation and signal. In a large population, she thus considers the consequences if a share $\kappa$ of the population used her strategy. This is not in general the same as an average of selfish and Kantian preferences. The two coincide when material payoffs are linear in others' actions, as with linear damages, and our marginal conditions then coincide with those of the two papers in the large-population limit under perfect information, where our aware consumers correspond to those in \citet{daube_moral_2016} and our unaware consumers to those in \citet{eichner_kantians_2022}. They differ when damages are convex: \emph{homo moralis} evaluates the damage at the load that would arise if a share $\kappa$ of the population used her strategy, whereas an average weighs the damage at the actual load against the damage that would arise if everybody acted as she does (Section~\ref{sec:ext:convex}). Our analysis also extends to imperfect information, which neither paper considers. Finally, we evaluate welfare by consumers' moral utility at realized outcomes, whereas \citet{daube_moral_2016} use material welfare and \citet{eichner_kantians_2022} lean towards it.

\citet{bierbrauer_morals_2026} study the political economy of a carbon tax whose revenue is rebated lump sum, in a population of neoclassical consumers, who maximize their own consumer surplus, and socially responsible consumers, who act as group-rule utilitarians: they choose the consumption rule that maximizes the welfare of the socially responsible if all of them follow it, taking the behavior of neoclassical consumers as given. With linear damages, their socially responsible consumers behave like aware \emph{homo moralis} consumers whose degree of morality equals the population share of socially responsible consumers, and their neoclassical consumers behave like \emph{homo economicus}. The foundations differ: for \emph{homo moralis}, the degree of morality is a preference parameter, and the thought experiment concerns all other consumers, whereas the moral weight of socially responsible consumers in \citet{bierbrauer_morals_2026} is determined by the composition of the population and only the welfare of the in-group counts. Moreover, their study lets the median voter rather than a benevolent planner set the tax, and they trace deviations from Pigovian taxation to distributive conflict between the groups and to moral pressure and resistance to it. 

Turning to the choice of policy instrument, \citet{herweg_how_2022} study how regulatory regimes affect morally motivated consumers. They show that a carbon tax complements consumers' voluntary efforts to reduce emissions, whereas cap-and-trade discourages them, since a binding cap fixes total emissions, and they conclude that taxation is more effective than cap-and-trade. Our analysis of quotas shares their offsetting mechanism: for consumers who are aware of the quota system, a sufficiently tight cap neutralizes morality (Proposition~\ref{prop:quota_aware}). Our conclusions nevertheless differ in two respects. First, a tax also neutralizes the morality of aware consumers when it is Pigovian. Moral motivation then complements a tax only if consumers are unaware of the tax-transfer system or the tax falls short of the Pigovian level. Second, we find no ranking of the instruments at their optima: with aware consumers, the best quota among those at least as stringent as the Pigovian quota reproduces the allocation under the Pigovian tax (Theorem~\ref{prop:quota_optimal}). Their ranking relies on an exogenous upper bound on the emissions price, reflecting the political opposition to efficient carbon prices that motivates their analysis.

\citet{kaufmann_understanding_2024} study rational consequentialist consumers who care about the externalities caused by their purchases. A binding cap makes a consumer's restraint ineffective, so such consumers behave selfishly, and for a planner who cannot implement the social optimum, a tax achieves a lower externality than a cap at the same permit fee. Our result for consumers aware of the quota system is the \emph{homo moralis} counterpart, with two differences: neutrality requires a cap at least as stringent as the Pigovian quota, and the best such quota reproduces the allocation under the Pigovian tax. Their survey evidence that about half of the respondents believe that their own consumption changes aggregate consumption one for one supports taking unaware consumers seriously.

\citet{besley_political_2023} study green transitions in a dynamic model in which consumption reflects not only prices and taxes but also green values. These values evolve endogenously, interact with the adoption of green technologies, and policy is set by politicians who cannot commit to future policies. Moral motivation thus takes the form of preferences that shift consumers' demand towards green goods. By contrast, we take degrees of morality as given, and moral concern operates through the Kantian thought experiment, so that its effect on behavior depends on the policy regime and on what consumers know about it.

Our approach also differs from other models of moral motivation. In models of warm glow \citep{andreoni_giving_1989,andreoni_impure_1990}, people derive utility from the prosocial act itself. \emph{Homo moralis} instead evaluates her strategy solely by the material consequences for herself that would follow if others used it too. Since the warm-glow payoff is attached to the act and does not depend on what happens to the tax revenue, a Pigovian tax does not neutralize it, in contrast to our neutrality result for aware consumers. \citet{brekke_economic_2003} endogenize warm glow as a self-image that depends on the distance between one's action and a morally ideal action, determined in a Kantian fashion as the action that would maximize social welfare if everybody took it. Policy then affects behavior also through its effect on the moral ideal, so that economic incentives may crowd out voluntary contributions. \emph{Homo moralis} preferences instead build Kantian morality into the evaluation of material payoffs, in a way that has an evolutionary foundation \citep{alger_homo_2013,alger_evolution_2016,alger_evolutionary_2019}.

In a field experiment, \citet{gneezy_fine_2000} found that a small fine for late pickup at Israeli day-care centers \emph{increased} late pickups, and that the increase persisted after the fine was withdrawn.\footnote{In their day-care setting the fine goes to the day-care center, not back to the parents as a group, so the restitution channel in our model does not map cleanly onto that experiment.} Our model cannot rationalize this finding. A fine that is the same for all units raises the consumption threshold by the full amount of the fine for unaware consumers and by the share $1-\kappa$ of it for aware consumers, so it cannot increase the fined behavior, and the model has no dynamics that could make an effect persist. Awareness does work as a restitution channel in our model: for an aware consumer, the returned revenue reduces the moral cost of a purchase by the share $\kappa$ of the fine. But the fine raises the price by its full amount, and the price effect dominates. \citet{ellingsen_model_2026} propose a model of duty-based motivation that distinguishes unconditional duties from conditional duties, which depend on the consequences of actions, and strict duties from liberal duties. The disutility from violating a duty is proportional to the externality caused, with strict and liberal violations weighted differently. In their Online Appendix, they show that their model can rationalize the finding of \citet{gneezy_fine_2000} through restitution: when the fine compensates those who bear the externality, it alleviates the guilt of dutiful parents, and a parent for whom the private cost of compliance exceeds the externality, and whose dutifulness lies in an intermediate range, becomes less punctual once the fine is introduced. We view this guilt mechanism, like the crowding-out mechanism in \citet{brekke_economic_2003}, as complementary to the \emph{homo moralis} mechanism we analyze.

\section{Concluding remarks}
\label{sec:discussion}

This paper has incorporated Kantian morality into a model of consumer
behavior and derived implications that modify standard policy prescriptions.
A general conclusion is that, under \emph{homo moralis} preferences, the
behavioral response to environmental policy is not merely a function of
prices and individual preferences, but also of the interplay between moral
motives and the perceived structure of the environmental tax system.
Policy-makers aiming to design effective environmental policy must therefore
consider not only what the price signal is, but also how that signal is
understood by a public desiring to do ``the right thing" to some extent.

In future work, it would be of interest to try to relax some of our
simplifying assumptions. In the main part of the analysis, we model binary unit-demand decisions, which is
appropriate for many small, discrete purchases but abstracts from income
effects and continuous consumption choices. 
In Section \ref{rem:continuous} we briefly consider continuous quantity choices for a single good category with quasi-linear utility. Extending the \emph{homo moralis} framework to a standard consumer budget set with multiple goods and income effects would be a natural and important next step. Another relevant extension would be to analyze mixed awareness more generally than in Section~\ref{sec:ext:awareness}, which considers perfectly informed consumers with a common degree of morality.

\bibliographystyle{jfe}
\bibliography{references.bib}

\newpage
\appendix{}
\noindent {\Huge \textbf{Appendix}}\bigskip

\begin{proof}[Derivation of equation \eqref{eq:W}]
Write $\Delta _{i}=Z_{i}-\tau \left( Z_{i},S_{i}\right) $ and $x_{i}^{\ast
}=x_{i}^{\ast }\left( V_{i},P_{i},S_{i}\right) $. Averaging $U_{i}^{\ast }$
over $i$,%
\begin{equation*}
W\left( \tau \right) =\frac{1}{n}\sum_{i}\mathbb{E}\left[ \left(
V_{i}-P_{i}-\tilde{\kappa}_{i}\Delta _{i}\right) x_{i}^{\ast }\right] -\frac{1}{%
n^{2}}\sum_{i}\left( 1-\kappa _{i}\right) \sum_{j\neq i}\mathbb{E}\left[
\Delta _{j}x_{j}^{\ast }\right] .
\end{equation*}%
Interchanging the order of summation in the last term,%
\begin{equation*}
\sum_{i}\left( 1-\kappa _{i}\right) \sum_{j\neq i}\mathbb{E}\left[
\Delta _{j}x_{j}^{\ast }\right] =\sum_{j}\mathbb{E}\left[ \Delta _{j}x_{j}^{\ast }\right]
\left( n\left( 1-\bar{\kappa}\right) -\left( 1-\kappa _{j}\right) \right) .
\end{equation*}%
The total coefficient on $\mathbb{E}\left[ \Delta _{i}x_{i}^{\ast }\right] $ in $%
W$ is therefore%
\begin{equation*}
-\frac{1}{n}\left[ \frac{1}{n}+\frac{n-1}{n}\kappa _{i}+\left( 1-\bar{\kappa%
}\right) -\frac{1-\kappa _{i}}{n}\right] =-\frac{1}{n}\left( 1+\kappa _{i}-%
\bar{\kappa}\right) ,
\end{equation*}%
so that%
\begin{equation*}
W\left( \tau \right) =\frac{1}{n}\sum_{i}\mathbb{E}\left[ \left(
V_{i}-P_{i}-\left( 1+\kappa _{i}-\bar{\kappa}\right) \left( Z_{i}-\tau
\left( Z_{i},S_{i}\right) \right) \right) x_{i}^{\ast }\right] .
\end{equation*}%
Since $P_{i}=\omega \left( Z_{i}\right) +\tau \left( Z_{i},S_{i}\right) $,
the integrand equals $V_{i}-\omega \left( Z_{i}\right) -Z_{i}-\left( \kappa
_{i}-\bar{\kappa}\right) \left( Z_{i}-\tau \left( Z_{i},S_{i}\right)
\right) $, and $x_{i}^{\ast }=1$ if and only if $V_{i}>A_{i}$. As $V_{i}$
is independent of $\left( Z_{i},S_{i}\right) $, integrating over $V_{i}$
conditional on $\left( Z_{i},S_{i}\right) $ yields \eqref{eq:W}.
\end{proof}

\bigskip

\bigskip

\begin{proof}[Proof of Proposition \protect\ref{thm:existence}]
Write $c=\omega \left( z\right) +z$ and $G\left( a\right) =\int_{a}^{\infty
}\left( v-c\right) f\left( v\right) dv$, the conditional welfare when
consumers buy if and only if their valuation exceeds $a$. Then $G^{\prime
}\left( a\right) =\left( c-a\right) f\left( a\right) $ for $a>0$, which is
positive for $a<c$ and negative for $a>c$, since $f>0$ on $\mathbb{R}_{+}$.
For $a\leq 0$ all consumers buy, so $G\left( a\right) =G\left( 0\right)
<G\left( c\right) $. Hence $G$ is uniquely maximized at $a=c=\omega \left(
z\right) +z$.

For $\alpha =0$, the threshold $a\left( z,s\right) =\omega \left( z\right)
+\tau \left( z,s\right) +\kappa \hat{z}_{s}$ is strictly increasing in $\tau
\left( z,s\right) $ and does not depend on the tax at other $\left(
z^{\prime },s^{\prime }\right) $. Hence $W\left( \tau \mid z,s\right) $ is
uniquely maximized at $\tau \left( z,s\right) =z-\kappa \hat{z}_{s}$.

For $\alpha =1$, pointwise maximization is not available, but it suffices to maximize $W\left( \tau \mid s\right) =\sum_{z}\Pr \left( Z=z\mid S=s\right) W\left( \tau \mid z,s\right) $ separately for each $s$. Recall that $\hat{r}_{s}=\mathbb{E}\left[ \tau \left( Z,s\right) \mid S=s\right] $, and define
\begin{equation*}
u\left( z,s\right) =\tau \left( z,s\right) -\kappa \hat{r}_{s},
\end{equation*}%
so that $a\left( z,s\right) =\omega \left( z\right) +u\left( z,s\right)
+\kappa \hat{z}_{s}$. Taking the expectation of $u\left( Z,s\right) $ given $%
S=s$ gives $\hat{u}_{s}=\left( 1-\kappa \right) \hat{r}_{s}$, hence $\tau
\left( z,s\right) =u\left( z,s\right) +\kappa \hat{u}_{s}/\left( 1-\kappa
\right) $. Since $\kappa <1$, the map $\tau \left( \cdot ,s\right) \mapsto
u\left( \cdot ,s\right) $ is a bijection for every $s$. In terms of $u$,%
\begin{equation*}
W\left( \tau \mid s\right) =\sum_{z}\Pr \left( Z=z\mid S=s\right)
\int_{\omega \left( z\right) +u\left( z,s\right) +\kappa \hat{z}%
_{s}}^{\infty }\left( v-\omega \left( z\right) -z\right) f\left( v\right)
dv,
\end{equation*}%
which depends on $u\left( \cdot ,s\right) $ only through its local value $%
u\left( z,s\right) $ in each term. By the first paragraph, each term is
uniquely maximized at $u\left( z,s\right) =z-\kappa \hat{z}_{s}$. Then $\hat{%
u}_{s}=\left( 1-\kappa \right) \hat{z}_{s}$, and hence $\tau \left(
z,s\right) =z-\kappa \hat{z}_{s}+\kappa \hat{z}_{s}=z$ for all $\left(
z,s\right) $. This is thus the unique welfare maximizer.
\end{proof}

{\normalsize \bigskip }

\begin{proof}[Proof of Theorem \protect\ref{P existence}]
Recall that $W\left( \tau \right) =\sum_{s}\Pr \left( S=s\right) W\left(
\tau \mid s\right) $, where%
\begin{equation*}
W\left( \tau \mid s\right) =\sum_{z^{\prime }}p_{z^{\prime }|s}W\left( \tau
\mid z^{\prime },s\right) ,\qquad p_{z^{\prime }|s}=\Pr \left( Z=z^{\prime
}\mid S=s\right) ,
\end{equation*}%
and that $W\left( \tau \mid s\right) $ depends on $\tau $ only through $\tau
\left( \cdot ,s\right) $. Fix $s$ and $z$ in the support of $Z\mid S=s$.
With $\alpha =1$, the threshold is%
\begin{equation*}
a\left( z^{\prime },s,\kappa \right) =\omega \left( z^{\prime }\right) +\tau
\left( z^{\prime },s\right) +\kappa \left( \hat{z}_{s}-\hat{r}_{s}\right) .
\end{equation*}%
Let%
\begin{equation*}
B\left( z^{\prime },s,\kappa \right) =a\left( z^{\prime },s,\kappa \right)
-\omega \left( z^{\prime }\right) -z^{\prime }-\left( \kappa -\bar{\kappa}%
\right) \left( z^{\prime }-\tau \left( z^{\prime },s\right) \right)
\end{equation*}%
denote the integrand in $W\left( \tau \mid z^{\prime },s\right) $ evaluated
at the threshold.  Since $\hat{r}_{s}=\sum_{z^{\prime }}p_{z^{\prime }|s}\tau \left( z^{\prime },s\right) $, we have%
\begin{equation*}
\frac{\partial a\left( z^{\prime },s,\kappa \right) }{\partial \tau \left(
z,s\right) }=\mathbf{1}\left\{ z^{\prime }=z\right\} -\kappa p_{z|s}.
\end{equation*}%
The integrand in $W\left( \tau \mid z^{\prime },s\right) $ depends on $\tau
\left( z,s\right) $ only if $z^{\prime }=z$. Applying Leibniz' rule term by
term, therefore,%
\begin{eqnarray*}
\frac{\partial W\left( \tau \mid s\right) }{\partial \tau \left( z,s\right) }
&=&p_{z|s}\mathbb{E}_{\kappa }\left[ \left( \kappa -\bar{\kappa}\right)
\left( 1-F\left( a\left( z,s,\kappa \right) \right) \right) \right] -p_{z|s}%
\mathbb{E}_{\kappa }\left[ B\left( z,s,\kappa \right) f\left( a\left(
z,s,\kappa \right) \right) \right] \\
&&+p_{z|s}\mathbb{E}_{\kappa }\left[ \kappa \sum_{z^{\prime }}p_{z^{\prime
}|s}B\left( z^{\prime },s,\kappa \right) f\left( a\left( z^{\prime
},s,\kappa \right) \right) \right] .
\end{eqnarray*}%
Dividing by $p_{z|s}>0$, and using that $\kappa $ is independent of $\left(
Z,S\right) $, the necessary first-order condition for an optimal choice of $%
\tau \left( z,s\right) $ is%
\begin{eqnarray}
0 &=&\mathbb{E}_{\kappa }\left[ \left( \kappa -\bar{\kappa}\right) \left(
1-F\left( a\left( z,s,\kappa \right) \right) \right) \right] -\mathbb{E}%
_{\kappa }\left[ B\left( z,s,\kappa \right) f\left( a\left( z,s,\kappa
\right) \right) \right]  \label{FOC:full} \\
&&+\mathbb{E}\left[ \kappa B\left( Z,s,\kappa \right) f\left( a\left(
Z,s,\kappa \right) \right) \mid S=s\right] ,  \notag
\end{eqnarray}%
where the last expectation is taken over $\left( Z,\kappa \right) $. For
perfectly informed consumers, where $Z\mid S=s$ is degenerate at $z=s$, %
\eqref{FOC:full} reduces to \eqref{FOC:heterotax} with $\alpha =1$.

We now show that $\tau \left( z,s\right) =z$ for all $\left( z,s\right) $
satisfies \eqref{FOC:full}. Under this tax, $\hat r_s =\hat{z}_{s}$, so the threshold reduces to $%
a\left( z,s,\kappa \right) =\omega \left( z\right) +z$ for all $\kappa $ and
all $\left( z,s\right) $. Consequently, $B\left( z,s,\kappa \right) =0$ for
all $\kappa $ and all $\left( z,s\right) $, so the second and third terms in %
\eqref{FOC:full} vanish. Since $a\left( z,s,\kappa \right) $ does not depend
on $\kappa $, the first term equals $\left( 1-F\left( \omega \left( z\right)
+z\right) \right) \mathbb{E}\left[ \kappa -\bar{\kappa}\right] =0$. Hence %
\eqref{FOC:full} is satisfied at every $\left( z,s\right) $.

\end{proof}

{\normalsize \bigskip }

\begin{lemma}
{\normalsize \label{lem:variance_bound} Let $\kappa $ be a random variable
whose support is contained in $[0,K]$ for some $K<2\sqrt{2}-2$, and let $%
L(\theta )=\log \mathbb{E}[e^{\theta \kappa }]$ be its cumulant generating function. Then $L^{\prime \prime }(\theta )<\left( 1-L^{\prime }(\theta )\right) ^{2}$ for all $\theta \in \mathbb{R}$. }
\end{lemma}

\begin{proof}
{\normalsize Let $M(\theta)=\mathbb{E}[e^{\theta\kappa}]$, so $%
L(\theta)=\log M(\theta)$. Direct differentiation gives
\begin{equation*}
m(\theta)=L^{\prime }(\theta)=\frac{\mathbb{E}[\kappa e^{\theta\kappa}]}{%
\mathbb{E}[e^{\theta\kappa}]}, \qquad \sigma^2(\theta)=L^{\prime \prime
}(\theta)=\frac{\mathbb{E}[\kappa^2 e^{\theta\kappa}]}{\mathbb{E}%
[e^{\theta\kappa}]}-m(\theta)^2.
\end{equation*}
Thus $m(\theta)$ and $\sigma^2(\theta)$ are the mean and variance of $\kappa$
under the probability measure $\mathbb{P}_\theta$ with density $%
e^{\theta\kappa}/M(\theta)$ relative to the distribution of $\kappa$. Since
this reweighting does not change the support, $\kappa$ takes values in $%
[0,K] $ under $\mathbb{P}_\theta$ as well. The Bhatia--Davis inequality,
applied to a random variable supported in $[0,K]$ with mean $m(\theta)$,
therefore yields
\begin{equation*}
\sigma^2(\theta)\leq m(\theta)\bigl(K-m(\theta)\bigr).
\end{equation*}
It suffices to show that $m(K-m)<(1-m)^2$ for all $m\in[0,K]$, or
equivalently $\rho(m)>0$ where
\begin{equation*}
\rho(m)=2m^2-(K+2)m+1.
\end{equation*}
The discriminant of $\rho$ is $(K+2)^2-8$, which is negative since $K<2\sqrt{%
2}-2$. Hence $\rho(m)>0$ for all $m\in\mathbb{R}$, and in particular for all
$m\in[0,K]$. }
\end{proof}

{\normalsize \bigskip }

\begin{proof}[Proof of Theorem \protect\ref%
{thm:unique_tax_aware_and_informed}]
{\normalsize With $\alpha =1$ and $S=Z$, the conditional welfare at environmental cost $z$ is
\begin{equation*}
W(t\mid z)=\mathbb{E}\left[ \int_{a(\kappa ,t)}^{\infty }\bigl(v-\omega
\left( z\right) -z-(\kappa -\bar{\kappa})(z-t)\bigr)f(v)\,dv\right] ,
\end{equation*}%
where $t=\tau (z)$ and $a(\kappa ,t)=\omega \left( z\right) +(1-\kappa
)t+\kappa z$. }

All thresholds are nonnegative, since $a\left( \kappa ,t\right) =\left( 1-\kappa \right) \left( \omega \left( z\right) +t\right) +\kappa \left( \omega \left( z\right) +z\right) $, where $\omega \left( z\right) +t\geq 0$ by assumption and $\omega \left( z\right) +z>0$. Hence $1-F\left( a\right) =e^{-a/\mu }$ and the memoryless identity below apply for every $\kappa $ in the support.

{\normalsize We first simplify the inner integral. Using $a-\omega \left(
z\right) -z=(1-\kappa )(t-z)$, the integrand can be rewritten as
\begin{equation*}
v-\omega \left( z\right) -z-(\kappa -\bar{\kappa})(z-t)=(v-a)+(1-\bar{\kappa}%
)(t-z).
\end{equation*}%
Hence
\begin{equation*}
\int_{a}^{\infty }\!\bigl(v-\omega \left( z\right) -z-(\kappa -\bar{\kappa}%
)(z-t)\bigr)f(v)\,dv=\int_{a}^{\infty }\!(v-a)f(v)\,dv+(1-\bar{\kappa})(t-z)%
\bigl(1-F(a)\bigr).
\end{equation*}%
The memoryless property of the exponential distribution gives $%
\int_{a}^{\infty }(v-a)f(v)\,dv=\mu \bigl(1-F(a)\bigr)$, so the inner
integral equals $\bigl[\mu +(1-\bar{\kappa})(t-z)\bigr]\bigl(1-F(a)\bigr)$.
Since the bracketed factor does not depend on $\kappa $, taking the
expectation over $\kappa $ yields the factorization
\begin{equation}
W(t\mid z)=\left[ (1-\bar{\kappa})(t-z)+\mu \right] \cdot N(t),
\label{eq:W_factored}
\end{equation}%
where $N(t)=\mathbb{E}\bigl[1-F\bigl(a(\kappa ,t)\bigr)\bigr]=e^{-(\omega
\left( z\right) +t)/\mu }\,M\!\left( \frac{t-z}{\mu }\right) >0$ and $%
M(\theta )=\mathbb{E}[e^{\theta \kappa }]$ is the moment generating function
of $\kappa $. }

{\normalsize Writing $h(t)=(1-\bar\kappa)(t-z)+\mu$ and $\theta=(t-z)/\mu$,
we have $h^{\prime }(t)=1-\bar\kappa$ and $N^{\prime
}(t)/N(t)=(m(\theta)-1)/\mu$, where $m(\theta)=M^{\prime }(\theta)/M(\theta)$%
. Since $L=\log M$, we have $m(\theta)=L^{\prime }(\theta)$, and a direct
calculation gives $L^{\prime \prime }(\theta )=M^{\prime \prime }(\theta )/M(\theta )-\left( M^{\prime }(\theta )/M(\theta )\right) ^{2}$. Write $%
\sigma^2(\theta)=L^{\prime \prime }(\theta)$. }

{\normalsize Define $g(\theta)=\frac{1}{1-m(\theta)}-\theta$. Then $%
g^{\prime }(\theta)=\frac{\sigma^2(\theta)}{(1-m(\theta))^2}-1<0$ by Lemma %
\ref{lem:variance_bound}, so $g$ is strictly decreasing, with $%
g(0)=1/(1-\bar\kappa)$. For $t$ with $h(t)>0$, dividing $W^{\prime
}(t)=h^{\prime }(t)N(t)+h(t)N^{\prime }(t)$ by the positive quantity $%
h(t)N(t)/\mu$ shows that the sign of $W^{\prime }(t)$ equals the sign of $%
g(\theta)-1/(1-\bar\kappa)$, where $\theta=(t-z)/\mu$. Hence $W^{\prime
}(t)>0$ for $t<z$ with $h(t)>0$, and $W^{\prime }(t)<0$ for $t>z$. On the
remaining region $h(t)\leq 0$ we have $W(t\mid z)=h(t)N(t)\leq 0<W(z\mid z)$%
, so the maximum cannot lie there. Therefore $t=z$ is the unique global
maximum of $W(\cdot\mid z)$. }
\end{proof}

{\normalsize \bigskip }

\begin{proof}[Proof of Proposition \protect\ref{prop:consumption_heteromoral}]
Write $P=\omega \left( Z\right) +\tau \left( Z,S\right) $ and $D=\hat{Z}-\alpha \hat{R}$, so that the random
threshold is $A=P+\kappa D$, where $\kappa $ is independent of $\left(
Z,S\right) $ with mean $\bar{\kappa}$. Let $w=\min_{z\in \mathcal{Z}}\omega
\left( z\right) $. Since $\tau \geq 0$, $D\geq 0$ and $\kappa \geq 0$, we
have $A\geq \omega \left( Z\right) \geq w$, and by assumption [M] the
function $1-F$ is strictly convex on $\left[ w,\infty \right) $.
Conditional on $\left( Z,S\right) $, the threshold is an affine function of
$\kappa $, so Jensen's inequality yields%
\begin{equation*}
\mathbb{E}\left[ 1-F\left( P+\kappa D\right) \mid Z,S\right] \geq 1-F\left(
P+\bar{\kappa}D\right) ,
\end{equation*}%
with strict inequality on the event $D>0$, since $\kappa $ is
non-degenerate and $1-F$ is strictly convex on the relevant range. Taking
expectations over $\left( Z,S\right) $, and using that $D>0$ with positive
probability, gives $C\left( \tau \right) >C_{\hom }\left( \tau \right) $.

For the environmental cost, multiply the conditional inequality by $Z\geq 0$
before taking expectations. On the event $D>0$ we have $\mathbb{E}\left[
Z\mid S\right] \geq D>0$, so $Z>0$ with positive conditional probability
given $S$, and the inequality remains strict after taking expectations.
Hence $E\left( \tau \right) >E_{\hom }\left( \tau \right) $.
\end{proof}

{\normalsize \bigskip }

\begin{proof}[Proof of Proposition \protect\ref{prop:info}]
Let $p$ denote the constant production cost, which is then also the consumer price. With no tax, the consumption thresholds under no, general and perfect information are $p+\kappa \bar{z}$, $p+\kappa \hat{Z}$ and $p+\kappa Z$, respectively, where $\hat{Z}=\mathbb{E}\left[ Z\mid S\right] $. Since $Z\geq 0$ and $\kappa \geq 0$, all three thresholds lie in $\left[ p,\infty \right) $, and on this set $1-F$ is strictly convex by assumption [M]. As $\kappa $ is independent of $\left( Z,S\right) $, we have $\mathbb{E}\left[ Z\mid S,\kappa \right] =\hat{Z}$ and $\mathbb{E}\left[ \hat{Z}\mid \kappa \right] =\bar{z}$. Jensen's inequality conditional on $\left( S,\kappa \right) $ gives%
\begin{equation*}
\mathbb{E}\left[ 1-F\left( p+\kappa Z\right) \mid S,\kappa \right] \geq 1-F\left( p+\kappa \hat{Z}\right) ,
\end{equation*}%
and Jensen's inequality conditional on $\kappa $ gives%
\begin{equation*}
\mathbb{E}\left[ 1-F\left( p+\kappa \hat{Z}\right) \mid \kappa \right] \geq 1-F\left( p+\kappa \bar{z}\right) .
\end{equation*}%
Taking expectations yields $C\left( 0,Z\right) \geq C\left( 0,S\right) \geq C\left( 0,\bar{z}\right) $.

Suppose now that $Z$ is non-degenerate and that the average degree of morality is positive, so that $\kappa >0$ with positive probability. On the event $\kappa >0$, the threshold $p+\kappa Z$ is non-degenerate conditional on $\kappa $, and strict convexity gives%
\begin{equation*}
\mathbb{E}\left[ 1-F\left( p+\kappa Z\right) \mid \kappa \right] >1-F\left( p+\kappa \bar{z}\right) .
\end{equation*}%
On the event $\kappa =0$, both sides equal $1-F\left( p\right) $. Taking expectations gives $C\left( 0,Z\right) >C\left( 0,\bar{z}\right) $.
\end{proof}

{\normalsize \bigskip }

\begin{proof}[Proof of Proposition \protect\ref{prop:awareness and morality}]
{\normalsize By Proposition \ref{thm:existence}, the optimal tax may be
written $\tau \left( Z,S\right) =Z-\left( 1-\alpha \right) \kappa \hat{Z}$.
The consumption threshold under the optimal tax is therefore%
\begin{eqnarray*}
A &=&\omega \left( Z\right) +\tau \left( Z,S\right) +\kappa \left( \hat{Z}%
-\alpha \hat R \right) \\
&=&\omega \left( Z\right) +Z-\left( 1-\alpha \right) \kappa \hat{Z}+\kappa
\left( \hat{Z}-\alpha \left( \hat{Z}-\left( 1-\alpha \right) \kappa \hat{Z}%
\right) \right) \\
&=&\omega \left( Z\right) +Z-\left( 1-\alpha \right) \kappa \hat{Z}+\kappa
\left( \hat{Z}-\alpha \hat{Z}\right) \\
&=&\omega \left( Z\right) +Z,
\end{eqnarray*}%
which does not depend on $\alpha $ (in the third line we used the fact that the product $(1-\alpha)\alpha=0$ when $\alpha=0$ and when $\alpha=1$). Hence, consumption is the same when $%
\alpha =0$ as when $\alpha =1$. The same is true for the environmental cost
of consumption. }
\end{proof}

\bigskip

\begin{proof}[Proof of Proposition \protect\ref{prop:quota_aware} and of the claim following it]
The function $Q$ is continuous and strictly decreasing, with $Q\left(\gamma \right) \rightarrow 0$ as $\gamma \rightarrow \infty $, so for $q\leq q^{\ast }=Q\left( 1\right) $ equation \eqref{tau q} has a unique solution $\gamma ^{\ast }\geq 1$. Let $x^{\ast }\left( v,p,s\right) = \mathbf{1}\left\{ v>p\right\} $, and suppose all consumers other than $i$ use $x^{\ast }$. At price $\gamma $ they buy if and only if $V>\omega \left( Z\right) +\gamma Z$, so their per capita emissions are $Q\left(\gamma \right) $.

Consider consumer $i$ with degree of morality $\kappa _{i}<1$ and an arbitrary strategy $x$. She evaluates $x$ in the hypothetical where a share $\kappa _{i}$ of the population uses $x$ and the rest use $x^{\ast }$, at a price $\tilde{\gamma}$ that clears the allowance market in that hypothetical. Write $\tilde{P}=\omega \left( Z\right) +\tilde{\gamma}Z$ for the hypothetical price, $Y=x(V,\tilde{P},S) $ for her decision at that price, and $c=\mathbb{E}\left[ ZY\right] $.
Hypothetical per capita emissions are%
\begin{equation*}
e=\kappa _{i}c+\left( 1-\kappa _{i}\right) Q( \tilde{\gamma}) ,
\end{equation*}%
and market clearing means either $\tilde{\gamma}>0$ and $e=q$, or $\tilde{\gamma}=0$ and $e\leq q$. In the hypothetical, every consumer experiences the environmental cost $e$ and receives the allowance revenue $\tilde{\gamma}e$, so her utility from $x$ is%
\begin{equation}
\tilde{U}_{i}\left( x\right) =\mathbb{E}\left[ ( V-\tilde{P}) Y\right] -( 1-\tilde{\gamma}) e.  \label{eq:quota_hyp}
\end{equation}%
For $x=x^{\ast }$ we have $c=Q( \tilde{\gamma}) $, so $\tilde{\gamma} = \gamma ^{\ast }$, $e=q$, $Y=\mathbf{1}\left\{ V>\omega \left( Z\right) + \gamma ^{\ast }Z\right\} $, and%
\begin{equation*}
\tilde{U}_{i}\left( x^{\ast }\right) =\mathbb{E}\left[ \left( V-\omega
\left( Z\right) -\gamma ^{\ast }Z\right) ^{+}\right] -\left( 1-\gamma
^{\ast }\right) q.
\end{equation*}%
We show that $\tilde{U}_{i}\left( x\right) \leq \tilde{U}_{i}\left(
x^{\ast }\right) $ for every $x$ and every market-clearing $\tilde{\gamma}$,
with equality only if $x( V,\omega \left( Z\right) +\gamma ^{\ast
}Z,S) =x^{\ast }( V,\omega \left( Z\right) +\gamma ^{\ast
}Z,S) $ almost surely. Throughout we use the pointwise bound%
\begin{equation}
\mathbb{E}\left[ \left( V-\omega \left( Z\right) -\gamma ^{\ast }Z\right) Y%
\right] \leq \mathbb{E}\left[ \left( V-\omega \left( Z\right) -\gamma
^{\ast }Z\right) ^{+}\right] ,  \label{eq:quota_pointwise}
\end{equation}%
which holds for any $\left\{ 0,1\right\} $-valued $Y$, with equality if
and only if $Y=\mathbf{1}\left\{ V>\omega \left( Z\right) +\gamma ^{\ast
}Z\right\} $ almost surely, since $V$ has a density.

\textit{Case 1: }$e=q$. Since $$\tilde{P}=\omega \left( Z\right) +\gamma
^{\ast }Z-\left( \gamma ^{\ast }-\tilde{\gamma}\right) Z,$$ we have $$\mathbb{%
E}[ ( V-\tilde{P}) Y] =\mathbb{E}\left[ \left( V-\omega
\left( Z\right) -\gamma ^{\ast }Z\right) Y\right] +\left( \gamma ^{\ast }-%
\tilde{\gamma}\right) c,$$ and by \eqref{eq:quota_pointwise},
\begin{equation*}
\tilde{U}_{i}\left( x\right) -\tilde{U}_{i}\left( x^{\ast }\right) \leq
\left( \gamma ^{\ast }-\tilde{\gamma}\right) c-( 1-\tilde{\gamma})
q+\left( 1-\gamma ^{\ast }\right) q=\left( \gamma ^{\ast }-\tilde{\gamma}%
\right) \left( c-q\right) .
\end{equation*}
From $e=q=Q\left( \gamma ^{\ast }\right) $, writing $Q\left( \gamma ^{\ast
}\right) =\kappa _{i}Q\left( \gamma ^{\ast }\right) +\left( 1-\kappa
_{i}\right) Q\left( \gamma ^{\ast }\right) $ and rearranging, we get $%
\kappa _{i}\left( c-q\right) =\left( 1-\kappa _{i}\right) \left( Q\left(
\gamma ^{\ast }\right) -Q( \tilde{\gamma}) \right) $.
Since $Q$ is strictly decreasing, $c-q$ has the sign of $\tilde{\gamma}-\gamma ^{\ast }$ when $%
\kappa _{i}\in \left( 0,1\right) $, and $\tilde{\gamma}=\gamma ^{\ast }$
when $\kappa _{i}=0$. In both cases $\left( \gamma ^{\ast }-\tilde{\gamma}%
\right) \left( c-q\right) \leq 0$.
If equality holds in the bound, then \eqref{eq:quota_pointwise} is tight, so $Y=\mathbf{1}\left\{ V>\omega
\left( Z\right) +\gamma ^{\ast }Z\right\} $ a.s. and hence $c=Q\left(
\gamma ^{\ast }\right) =q$.
Since $\kappa _{i}<1$, the clearing condition $\kappa _{i}c+\left(
1-\kappa _{i}\right) Q( \tilde{\gamma}) =q$ with $c=q$ then gives $%
Q( \tilde{\gamma}) =q$, so $\tilde{\gamma}=\gamma ^{\ast }$ and $x$ agrees a.s. with $x^{\ast }$ at the equilibrium price.

\textit{Case 2: }$\tilde{\gamma}=0$ and $e<q$. Then $\kappa _{i}>0$, since
$Q\left( 0\right) \geq Q\left( \gamma ^{\ast }\right) =q$, and $\kappa
_{i}c<q-\left( 1-\kappa _{i}\right) Q\left( 0\right) \leq \kappa _{i}q$
gives $c<q$. Now $\tilde{P}=\omega \left( Z\right) $ and $( 1-\tilde{%
\gamma}) e=e$, so by \eqref{eq:quota_hyp}%
\begin{equation*}
\tilde{U}_{i}\left( x\right) =\mathbb{E}\left[ \left( V-\omega \left(
Z\right) \right) Y\right] -\kappa _{i}c-\left( 1-\kappa _{i}\right) Q\left(
0\right) .
\end{equation*}%
Since $\omega \left( Z\right) =\omega \left( Z\right) +\gamma ^{\ast
}Z-\gamma ^{\ast }Z$, we have $\mathbb{E}\left[ \left( V-\omega \left(
Z\right) \right) Y\right] =\mathbb{E}\left[ \left( V-\omega \left( Z\right)
-\gamma ^{\ast }Z\right) Y\right] +\gamma ^{\ast }c$, and %
\eqref{eq:quota_pointwise} gives%
\begin{equation*}
\tilde{U}_{i}\left( x\right) \leq \mathbb{E}\left[ \left( V-\omega \left(
Z\right) -\gamma ^{\ast }Z\right) ^{+}\right] +\gamma ^{\ast }c-\kappa
_{i}c-\left( 1-\kappa _{i}\right) Q\left( 0\right) .
\end{equation*}%
Subtracting $\tilde{U}_{i}\left( x^{\ast }\right) =\mathbb{E}\left[ \left(
V-\omega \left( Z\right) -\gamma ^{\ast }Z\right) ^{+}\right] -\left(
1-\gamma ^{\ast }\right) q$,%
\begin{equation*}
\tilde{U}_{i}\left( x\right) -\tilde{U}_{i}\left( x^{\ast }\right) \leq
\left( \gamma ^{\ast }-\kappa _{i}\right) c-\left( 1-\kappa _{i}\right)
Q\left( 0\right) +\left( 1-\gamma ^{\ast }\right) q.
\end{equation*}%
Using $Q\left( 0\right) \geq q$ in the middle term,%
\begin{eqnarray*}
\tilde{U}_{i}\left( x\right) -\tilde{U}_{i}\left( x^{\ast }\right) &\leq
\left( \gamma ^{\ast }-\kappa _{i}\right) c-\left( 1-\kappa _{i}\right)
q+\left( 1-\gamma ^{\ast }\right) q \\
&=\left( \gamma ^{\ast }-\kappa
_{i}\right) c-\left( \gamma ^{\ast }-\kappa _{i}\right) q\\
&=\left( \gamma
^{\ast }-\kappa _{i}\right) \left( c-q\right) ,
\end{eqnarray*}%
since
$-\left( 1-\kappa _{i}\right) +\left( 1-\gamma ^{\ast }\right)
=-\left( \gamma ^{\ast }-\kappa _{i}\right) $.
Finally, $c<q$ and $\gamma
^{\ast }\geq 1>\kappa _{i}$ give $\left( \gamma ^{\ast }-\kappa
_{i}\right) \left( c-q\right) <0$.

Hence $x^{\ast }$ is a.s. the unique best response for every consumer when
all others use $x^{\ast }$, and $x^{\ast }$ together with the price $\gamma
^{\ast }$ solving \eqref{tau q} is an equilibrium. This proves the proposition.

For the claim following the proposition, let $q>q^{\ast }$, so that $\gamma ^{\ast }<1$, and
consider a perfectly informed consumer using $x\left( v,p,z\right) =\mathbf{%
1}\left\{ v>p+z\right\} $. At price zero this gives $Y=\mathbf{1}\left\{
V>\omega \left( Z\right) +Z\right\} $ and $c=q^{\ast }$, so hypothetical
emissions are $e=\kappa _{i}q^{\ast }+\left( 1-\kappa _{i}\right) Q\left(
0\right) $, which is below $q$ whenever $\kappa _{i}>\left( Q\left(
0\right) -q\right) /\left( Q\left( 0\right) -q^{\ast }\right) $, a
threshold strictly below $1$ since $q>q^{\ast }$. For such $\kappa _{i}$
the market clears at $\tilde{\gamma}=0$, and by \eqref{eq:quota_hyp}%
\begin{equation*}
\tilde{U}_{i}\left( x\right) =\mathbb{E}\left[ \left( V-\omega \left(
Z\right) \right) \mathbf{1}\left\{ V>\omega \left( Z\right) +Z\right\} %
\right] -\kappa _{i}q^{\ast }-\left( 1-\kappa _{i}\right) Q\left( 0\right) .
\end{equation*}%
As $\kappa _{i}\rightarrow 1$ this converges to $\mathbb{E}\left[ \left(
V-\omega \left( Z\right) -Z\right) ^{+}\right] $, whereas $$\tilde{U}%
_{i}\left( x^{\ast }\right) =\mathbb{E}\left[ \left( V-\omega \left(
Z\right) -Z\right) \mathbf{1}\left\{ V>\omega \left( Z\right) +\gamma
^{\ast }Z\right\} \right] $$ does not depend on $\kappa _{i}$ and is
strictly smaller since $\gamma ^{\ast }<1$. Hence $\tilde{U}_{i}\left(
x\right) >\tilde{U}_{i}\left( x^{\ast }\right) $ for $\kappa _{i}$
sufficiently close to $1$.
\end{proof}

{\normalsize \bigskip }

\begin{proof}[Proof of Theorem \protect\ref{prop:quota_optimal}]
By Proposition \ref{prop:quota_aware}, for any quota $q\leq q^{\ast }$ the allowance price is the unique solution $\gamma \geq 1$ to \eqref{tau q}, and every consumer buys if and only if $V>\omega \left( Z\right) +\gamma Z$, irrespective of her morality and signal. Since $Q$ is strictly decreasing, quotas $q\leq q^{\ast }$ correspond one-to-one to allowance prices $\gamma \geq 1$, and it suffices to maximize welfare over $\gamma
\geq 1$.

The payment for allowances acts as the tax $\tau \left( Z\right) =\gamma Z$ with revenue returned lump-sum,  so per capita expected utility at realized outcomes is given by the large-population welfare function with $\tau \left(
Z\right) =\gamma Z$ and threshold $A=\omega \left( Z\right) +\gamma Z$:%
\begin{equation*}
W\left( \gamma \right) =\mathbb{E}\left[ \int_{\omega \left( Z\right)
+\gamma Z}^{\infty }\left( v-\omega \left( Z\right) -Z-\left( \kappa -\bar{%
\kappa}\right) \left( 1-\gamma \right) Z\right) f\left( v\right) dv\right] .
\end{equation*}%
Since $\kappa $ is independent of $Z$ and $\mathbb{E}\left[ \kappa -\bar{%
\kappa}\right] =0$, while neither the threshold nor the remaining integrand
depends on $\kappa $, the last term in the integrand contributes zero, and%
\begin{equation*}
W\left( \gamma \right) =\mathbb{E}\left[ \int_{\omega \left( Z\right)
+\gamma Z}^{\infty }\left( v-\omega \left( Z\right) -Z\right) f\left(
v\right) dv\right] .
\end{equation*}%
By Leibniz' rule,%
\begin{equation*}
W^{\prime }\left( \gamma \right) =-\mathbb{E}\left[ Z\left( \gamma
-1\right) Zf\left( \omega \left( Z\right) +\gamma Z\right) \right] =-\left(
\gamma -1\right) \mathbb{E}\left[ Z^{2}f\left( \omega \left( Z\right)
+\gamma Z\right) \right] ,
\end{equation*}%
which is negative for $\gamma >1$, since $f>0$ and $Z>0$ with positive
probability. Hence $\gamma =1$ is the unique welfare-maximizing allowance
price among $\gamma \geq 1$, and the corresponding quota is $q^{\ast
}=Q\left( 1\right) $ by \eqref{tau q}. At $\gamma =1$ the consumption
threshold is $\omega \left( Z\right) +Z$, the same as under Pigovian
taxation.
\end{proof}

\bigskip

\begin{proof}[Proof of Lemma \protect\ref{lem:convex_opt}]
Fix $i$, $\bar{\ell}$ and $\bar{r}$, write $\kappa =\kappa _{i}$, and let $\bar{X}$ be the convex set of strategies taking values in $\left[ 0,1\right] $. Since $\ell\left( x\right) $ and $r\left( x\right) $ are linear in $x$ and $\psi$ is convex, $\tilde{U}_{i}$ is concave on $\bar{X}$.

For $x^{\ast },x\in \bar{X}$ and $t\in \left( 0,1\right] $, concavity gives
\begin{equation*}
\tilde{U}_{i}\left( x\right) -\tilde{U}_{i}\left( x^{\ast }\right) \leq \frac{\tilde{U}_{i}\left( x^{\ast }+t\left( x-x^{\ast }\right) \right) -\tilde{U}_{i}\left( x^{\ast }\right) }{t}.
\end{equation*}
As $t\downarrow 0$, the right-hand side converges to the directional derivative, since $\psi$ is continuously differentiable. Replacing $Z$ and $\tau \left( Z\right) $ by their conditional expectations $\hat{Z}$ and $\hat{R}$ given $S$, as in the derivation of \eqref{U cond}, the directional derivative is $\mathbb{E}\left[ g\left( x-x^{\ast }\right) \right] $, where
\begin{equation*}
g=V-P-\kappa \left( m\hat{Z}-\alpha \hat{R}\right) ,\qquad m=\psi^{\prime }\left( \kappa \ell\left( x^{\ast }\right) +\left( 1-\kappa \right) \bar{\ell}\right) .
\end{equation*}
Since $\hat{Z}$ and $\hat{R}$ are functions of $S$, and $V$ has a density and is independent of $\left( P,S\right) $, we have $g\neq 0$ almost surely.

\textit{Necessity.} Let $x^{\ast }$ be optimal. Then $\mathbb{E}\left[ g\left( x-x^{\ast }\right) \right] \leq 0$ for all $x\in \bar{X}$. Take $x=\mathbf{1}\left\{ g>0\right\} $. Since $x^{\ast }$ takes values in $\left[ 0,1\right] $, the integrand $g\left( \mathbf{1}\left\{ g>0\right\} -x^{\ast }\right) $ is nonnegative, so it vanishes almost surely. As $g\neq 0$ almost surely, $x^{\ast }=\mathbf{1}\left\{ g>0\right\} =x^{(m)}$ almost surely, and hence $\ell\left( x^{\ast }\right) =\ell^{(m)}$. Thus $m$ solves $m=\Phi \left( m\right) $, where $\Phi \left( m\right) =\psi^{\prime }\left( \kappa \ell^{(m)} +\left( 1-\kappa \right) \bar{\ell}\right) $.

\textit{Sufficiency.} Let $m$ solve $m=\Phi \left( m\right) $ and $x^{\ast }=x^{(m)}$. Then $x^{\ast }=\mathbf{1}\left\{ g>0\right\} $, so $g\left( x-x^{\ast }\right) \leq 0$ pointwise for every $x\in \bar{X}$. Hence the directional derivative is nonpositive, and $\tilde{U}_{i}\left( x\right) \leq \tilde{U}_{i}\left( x^{\ast }\right) $.

\textit{The fixed point.} The function $\ell^{(m)} =\mathbb{E}\left[ Z\mathbf{1}\left\{ V>P+\kappa \left( m\hat{Z}-\alpha \hat{R}\right) \right\} \right] $ is continuous in $m$, by dominated convergence, since $V$ has a density and $Z$ has finite support. It is decreasing in $m$, since $\hat{Z}\geq 0$. As $\psi^{\prime }$ is continuous, nonnegative and increasing, $\Phi $ is continuous, nonnegative and decreasing. Hence $m-\Phi \left( m\right) $ is continuous and strictly increasing, nonpositive at $m=0$ and unbounded above, so $m=\Phi \left( m\right) $ has exactly one solution $m_{i}$.

By sufficiency, $x^{(m_{i})}$ is optimal. By necessity, every optimal strategy equals $x^{(m)}$ almost surely for a solution $m$ of $m=\Phi \left( m\right) $, that is, for $m=m_{i}$. If $\kappa =0$, the threshold is zero for every $m$ and the conclusion holds trivially.
\end{proof}

\bigskip

\begin{proof}[Proof of Proposition \protect\ref{prop:convex_pigou}]
Let the actual load be $\bar{\ell}=\ell^{\ast }$. Under the tax $\tau \left( z\right) =\psi^{\prime }\left( \ell^{\ast }\right) z$, linearity of conditional expectation gives $\hat{R}_{i}=\mathbb{E}\left[ \tau \left( Z_{i}\right) \mid S_{i}\right] =\psi^{\prime }\left( \ell^{\ast }\right) \hat{Z}_{i}$, whatever the signal. For $m=\psi^{\prime }\left( \ell^{\ast }\right) $, the threshold in \eqref{eq:x convex} is therefore $\kappa _{i}\left( m\hat{Z}_{i}-\psi^{\prime }\left( \ell^{\ast }\right) \hat{Z}_{i}\right) =0$, so $x_{i}^{(m)}=\mathbf{1}\left\{ V_{i}>P_{i}\right\} =x^{\ast }$, since $P_{i}=\omega \left( Z_{i}\right) +\psi^{\prime }\left( \ell^{\ast }\right) Z_{i}$. Hence $\ell_{i}^{(m)}=\ell\left( x^{\ast }\right) =\ell^{\ast }$ and $\kappa _{i}\ell_{i}^{(m)}+\left( 1-\kappa _{i}\right) \bar{\ell}=\ell^{\ast }$, so $m=\psi^{\prime }\left( \ell^{\ast }\right) $ solves the fixed-point equation in Lemma \ref{lem:convex_opt}. By uniqueness of that solution, $x^{\ast }$ is almost surely consumer $i$'s unique best response, for every $\kappa _{i}$ and every signal structure. If all consumers use $x^{\ast }$, the actual load is $\ell\left( x^{\ast }\right) =\ell^{\ast }$, so this is an equilibrium, and by uniqueness of equilibrium it is the only one. The induced allocation is $x^{\ast }$, the first best. Finally, since all consumers use the same strategy, the counterfactual load and revenue in \eqref{eq:U convex} equal the actual ones, so each consumer's moral utility equals her material utility.
\end{proof}

\bigskip

\begin{proof}[Proof of Proposition \protect\ref{prop:mixed_awareness}]
Fix $z>0$ and write $t=\tau \left( z\right) $. Aware consumers face the
threshold $A_{1}=\omega \left( z\right) +\left( 1-\kappa \right) t+\kappa z$
and unaware consumers the threshold $A_{0}=\omega \left( z\right) +t+\kappa
z$. Since morality is homogeneous, the conditional welfare is%
\begin{equation*}
W\left( t\mid z\right) =\phi \int_{A_{1}}^{\infty }\left( v-\omega \left(
z\right) -z\right) f\left( v\right) dv+\left( 1-\phi \right)
\int_{A_{0}}^{\infty }\left( v-\omega \left( z\right) -z\right) f\left(
v\right) dv.
\end{equation*}%
By Leibniz' rule,
\begin{equation*}
W^{\prime }\left( t\mid z\right) =-\phi \left( 1-\kappa \right) ^{2}\left(
t-z\right) f\left( A_{1}\right) -\left( 1-\phi \right) \left( t-\left(
1-\kappa \right) z\right) f\left( A_{0}\right) .
\end{equation*}%
For $t\leq \left( 1-\kappa \right) z$ both terms are nonnegative. If
$A_{1}>0$, the first term is positive. If $A_{1}\leq 0$, then $t<0$ and
$A_{0}=A_{1}+\kappa t<0$, so both terms vanish since $f=0$ on $\left( -\infty
,0\right) $. Hence $W\left( \cdot \mid z\right) $ is nondecreasing on $\left(
-\infty ,\left( 1-\kappa \right) z\right] $ and strictly increasing on the part
of this interval where $A_{1}>0$, which contains a left neighborhood of
$\left( 1-\kappa \right) z$. For $t\geq z$ both terms are nonpositive and the
second is negative, since $t-\left( 1-\kappa \right) z\geq \kappa z>0$ and
$A_{0}>0$. Hence every maximizer lies in $\left( \left( 1-\kappa \right)
z,z\right) $, and a maximizer exists since $W\left( \cdot \mid z\right) $ is
continuous on $\left[ \left( 1-\kappa \right) z,z\right] $.

Finally, $\int_{A}^{\infty }\left( v-\omega \left( z\right) -z\right)
f\left( v\right) dv$ is uniquely maximized over $A$ at $A=\omega \left(
z\right) +z$, and $A_{1}=A_{0}=\omega \left( z\right) +z$ requires $t=z$
and $t=\left( 1-\kappa \right) z$ simultaneously, which is impossible for $%
\kappa z>0$. Hence $W\left( t\mid z\right) $ is strictly below the
first-best value for every $t$, whereas by Proposition \ref{prop:awareness
and morality} the first-best value is attained when $\phi \in \left\{
0,1\right\} $.
\end{proof}

\end{document}